\documentclass[reqno,10pt,a4paper,dvips]{amsart}

\usepackage{amssymb,times,mathptmx,cite,eucal,psfrag,array,setspace,geometry}
\usepackage[dvips]{graphicx}

\numberwithin{equation}{section}

\newcolumntype{C}{>{$}c<{$}} 

\allowdisplaybreaks

\newcommand{\alg}[1]{\mathfrak{#1}}

\newcommand{\func}[2]{#1 \left( #2 \right)}

\newcommand{\brac}[1]{\left( #1 \right)}
\newcommand{\sqbrac}[1]{\left[ #1 \right]}

\newcommand{\ZZ}{\mathbb{Z}}

\newcommand{\ket}[1]{\bigl\lvert #1 \bigr\rangle}
\newcommand{\braket}[2]{\bigl\langle #1 \bigr\rvert \bigl. #2 \bigr\rangle}
\newcommand{\bracket}[3]{\bigl\langle #1 \bigr\rvert #2 \bigl\lvert #3 \bigr\rangle} 
\newcommand{\corrfn}[1]{\bigl\langle #1 \bigr\rangle}

\newcommand{\normord}[1]{{} : #1 : {}} 

\newcommand{\VerMod}[1]{\mathcal{V}_{#1}}
\newcommand{\IrrMod}[1]{\mathcal{L}_{#1}}
\newcommand{\IndMod}[1]{\mathcal{M}_{#1}}
\newcommand{\LogMod}[1]{\mathcal{I}_{#1}}
\newcommand{\MegaMod}[1]{\mathcal{J}_{#1}}

\newcommand{\fuse}{\times_{\! f}}

\newcommand{\dses}[3]{0 \longrightarrow #1 \longrightarrow #2 \longrightarrow #3 \rightarrow 0}

\newcommand{\eqnref}[1]{Equation~(\ref{#1})}

\newcommand{\secref}[1]{Section~\ref{#1}}

\newcommand{\figref}[1]{Figure~\ref{#1}}
\newcommand{\figsref}[2]{Figures~\ref{#1} and \ref{#2}}
\newcommand{\tabref}[1]{Table~\ref{#1}}

\newcommand{\cft}{conformal field theory}
\newcommand{\cfts}{conformal field theories}
\newcommand{\bcft}{boundary conformal field theory}
\newcommand{\bcfts}{boundary conformal field theories}
\newcommand{\lcft}{logarithmic conformal field theory}
\newcommand{\lcfts}{logarithmic conformal field theories}
\newcommand{\sle}{stochastic Loewner evolution}
\newcommand{\ope}{operator product expansion}
\newcommand{\opes}{operator product expansions}
\newcommand{\hws}{highest weight state}
\newcommand{\hwss}{highest weight states}
\newcommand{\hwm}{highest weight module}
\newcommand{\hwms}{highest weight modules}
\newcommand{\hwsm}{highest weight submodule}
\newcommand{\hwsms}{highest weight submodules}
\newcommand{\ode}{ordinary differential equation}

\newcommand{\pde}{partial differential equation}
\newcommand{\pdes}{partial differential equations}

\DeclareMathOperator{\id}{id}

\begin{document}

\title[Percolation and Watts' Crossing Probability]{On the Percolation BCFT and the Crossing Probability of Watts}

\author[D Ridout]{David Ridout}

\address[David Ridout]{
Theory Group, DESY \\
Notkestra\ss{}e 85 \\
D-22603, Hamburg, Germany
}

\email{dridout@mail.desy.de}

\thanks{\today \\ Part of this work was supported by the Marie Curie Excellence Grant MEXT-CT-2006-042695.}

\begin{abstract}
The logarithmic conformal field theory describing critical percolation is further explored using Watts' determination of the probability that there exists a cluster connecting both horizontal and vertical edges.  The boundary condition changing operator which governs Watts' computation is identified with a primary field which does not fit naturally within the extended Kac table.  Instead a ``shifted'' extended Kac table is shown to be relevant.  Augmenting the previously known logarithmic theory based on Cardy's crossing probability by this field, a larger theory is obtained, in which new classes of indecomposable rank-$2$ modules are present.  No rank-$3$ Jordan cells are yet observed.  A highly non-trivial check of the identification of Watts' field is that no Gurarie-Ludwig-type inconsistencies are observed in this augmentation.  The article concludes with an extended discussion of various topics related to extending these results including projectivity, boundary sectors and inconsistency loopholes.
\end{abstract}

\maketitle

\onehalfspacing

\section{Introduction} \label{secIntro}

There has recently been a surge of interest concerning the exploration of the mathematical structure of the \cft{} describing percolation at its critical point.  Percolation itself refers to a collection of closely related problems in probability theory \cite{KesPer82,GriPer89}, but these problems exhibit behaviour analogous to that of phase transitions in macroscopic media \cite{LanCon94,CarLec01}.  The interest from the physics community then stems from regarding percolation as a collection of relatively simple statistical models with which one can test predictions such as conformal invariance at criticality and universality.

Perhaps the most celebrated confirmation of these predictions is Cardy's derivation of the horizontal crossing probability in the continuum limit \cite{CarCri92}.  This derivation assumes that the limit is conformally invariant, and relies upon standard \cft{} techniques \cite{DiFCon97}.  One considers a fixed rectangular subdomain of a square lattice, and considers random configurations in which each edge (bond) of the lattice is chosen to be open or closed with probability $p$ or $1-p$ respectively.  A fundamental question of percolation is then to calculate the probability that such a random configuration will contain a cluster of bonds connecting one of the vertical sides of the rectangle to the other.  In the continuum limit where the lattice spacing tends to zero (but the rectangular domain remains fixed), this crossing probability is only interesting (neither zero nor one) for a critical $p = p_c$, and Cardy's work gives this limiting crossing probability as a function of the aspect ratio of the rectangular domain.  The agreement with the numerical simulations of \cite{LanUni92} is impressive, and is generally agreed to provide a striking confirmation of the conformal invariance of statistical models at criticality.

It is worth noting that results such as Cardy's have encouraged probability theorists to find mathematically rigorous derivations of crossing probability formulae.  In particular, Cardy's result is now a theorem \cite{SmiCri01,LawValI01}.  The toolbox which has led to these successes is known collectively as \sle{} \cite{SchSca00}, and these last ten years have seen a rapid development in its understanding and application to statistical models.  In particular, one can now begin to ask questions regarding the precise relation between the points of view afforded by the \sle{} and \cft{} descriptions \cite{BauCon04,KytFro08}.

Here, we wish to restrict ourselves to the \cft{} description of critical percolation.  Much of the interest in this \cft{} stems from the simple realisation that it must be logarithmic (see \cite{CarLog99} for an early statement to this effect).  Indeed, it is almost universally agreed that critical percolation corresponds to a theory with vanishing central charge $c$ (though \cite{FloNot06} offers a dissenting opinion).  A standard argument \cite{FeiAnn92,RidPer07} then proves that any $c=0$ \cft{} built from \emph{irreducible} Virasoro modules is trivial\footnote{As one might expect, this ``irreducibility implies triviality'' argument generalises to affine Kac-Moody algebras when $c=0$.  However, the same is not true for affine Kac-Moody superalgebras.  For example, the irreducible vacuum $\func{\widehat{\alg{gl}}}{1 \mid 1}$-module (which always has $c=0$) is decidedly non-trivial.  However, it does (necessarily) decompose into Virasoro modules which are not irreducible, but merely indecomposable.}.  The alternative --- that the theory is built from reducible but \emph{indecomposable} Virasoro modules --- leads to so-called \lcft{} \cite{RozQua92,GurLog93}.

Intertwined with this story is that of the $c \rightarrow 0$ ``catastrophe'' \cite{CarLog99,GurCon02,CarStr02,KogStr02,GurCon04}.  Here, one asks what happens to the \ope{} of a primary field and its conjugate when $c \rightarrow 0$.  The standard form of this expansion shows that the coefficient of the energy-momentum tensor $\func{T}{z}$ diverges unless the dimension of the primary field also tends to $0$.  Resolving this issue involves modifying the \ope{} by adding ``partner fields'', and one quickly finds that this modification also leads to \lcft{}.  We mention that the derivation of the standard \ope{} between a primary field and its conjugate breaks down at $c=0$ \cite{RidMin07} (because the energy-momentum tensor is null), so the ``catastrophe'' alluded to above is merely an expression of the subtlety involved in making sense of limits such as $c \rightarrow 0$.  Exactly the same problem occurs with the $\func{\partial^2 T}{z}$ and $\normord{\func{T}{z} \func{T}{z}}$ terms as $c \rightarrow \tfrac{-22}{5}$, unless the primary field dimension tends to $0$ or $\tfrac{-1}{5}$ \cite{ZamInf85,RidMin07}.

The stage is now set for studying critical percolation via investigating $c=0$ \lcfts{}.  However, the number of such theories is probably infinite:  Aside from the theories describing percolation and the other $c=0$ statistical model, self-avoiding walks, there are theories constructed (in varying degrees) from affine Kac-Moody algebras \cite{CauLog97,NicLog01} and superalgebras \cite{RozQua92,BerQua01,LudFre00,SalGL106,SalSU207,CreBra08}.  Indeed, one expects to be able to construct \lcfts{} for each of the superalgebras $\func{\widehat{\alg{gl}}}{n \mid n}$, $\func{\widehat{\alg{sl}}}{n+1 \mid n}$, $\func{\widehat{\alg{osp}}}{2n \mid 2n}$ and $\func{\widehat{\alg{osp}}}{2n+1 \mid 2n}$, all of which will have vanishing central charge.  It is therefore of utmost importance to be clear as to how one \emph{identifies} a given $c=0$ \lcft{} as describing critical percolation.  In other words, it is insufficient to rely on mathematical consistency alone; there must be some physical input to the theory which selects the correct choice.

This said, there has recently been much progress made in identifying the percolation \cft{}.  One can isolate several different approaches including free field methods \cite{FjeLog02,FeiMod06,FeiLog06}, lattice model constructions \cite{ReaExa01,PeaLog06,ReaAss07,RasFus07,RasWEx08} and fusion \cite{EbeVir06,SimPer07,RidPer07,RidLog07}.  Each approach has its own advantages and disadvantages, and each seems to produce results which agree (broadly speaking) when they can be compared.  One criticism that can be levelled at much of the \lcft{} literature however is that it restricts consideration to the chiral sector.  Whilst this is natural in \cft{} proper, where the modular invariants of rational theories enjoy a simple factorisation property, it is difficult to justify in a \lcft{}.  Indeed, the few examples (see \cite{GabLoc99} for the first) in which modular invariants have been constructed show that this familiar factorisation property is absent in logarithmic theories.  In any case, the explicit form of the correlation functions in such theories makes it clear that na\"{\i}ve factorisation cannot suffice.

One situation in which restricting attention to chiral matters is nevertheless justified is when one is considering the boundary sectors of the theory.  It is clear from any discussion of crossing probabilities that percolation must be described by a \bcft{}, and indeed, this is the structure that Cardy exploited in his derivation of the horizontal crossing probability \cite{CarCri92}.  Generally in \cft, one is taught to understand the theory in the bulk first, as much of the boundary theory is then deducible from that of the bulk \cite{CarEff86,RunBou99}.  Recently however, it has been realised \cite{FucTFTI02,FjeUni06,GabFro08} that the reverse is also true, and that extracting the bulk from the boundary may be ``cleaner'' in some sense.

A further advantage of starting with the boundary sector of the percolation \cft{} is that one has well-known (and numerically tested!) results on which to base the theory.  This was the motivation for the construction detailed in \cite{RidPer07}; Cardy's crossing probability formula was used as the initial \emph{physical} datum for a detailed exploration of the boundary sector of the percolation \cft{}.  From this alone, the structures confirming the logarithmic nature of the theory were generated.

We will review this construction shortly, along with the assertion that the operator content of a \lcft{} is strongly constrained by the requirements of conformal invariance \cite{GurCon04,RidLog07}.  What is relevant for now, and should be clear, is that this construction is only the beginning of the story.  There are other observables of critical percolation with which one can play a similar game, in particular, there is the probability of there being a percolation cluster connecting both the vertical \emph{and} horizontal sides of the rectangular domain.  This crossing probability is the subject of an article of Watts \cite{WatCro96}, in which he derives a formula which again interpolates the numerical data beautifully \cite{LanUni92}.  We are therefore led to reflect on whether Watts' result is incorporated in the \lcft{} proposed in \cite{RidPer07}.  We will argue below that it is not, and the consequences of this realisation constitute the core of this article.

Let us take this opportunity to introduce some useful notation \cite{RidPer07,RidLog07}.  First, we let $\VerMod{r,s}$ ($r,s \in \ZZ_+$) denote the Verma module for the Virasoro algebra $\alg{Vir}$ (with vanishing central charge) whose \hws{} has conformal dimension
\begin{equation} \label{eqnConfDim}
h_{r,s} = \frac{\brac{3r-2s}^2-1}{24}.
\end{equation}
These dimensions are conveniently arranged in a semi-infinite table which we shall refer to as the $c=0$ extended Kac table.  A part of this table is presented in \tabref{tabExtKacc=0}.  The Verma module $\VerMod{r,s}$ has a singular vector at grade $rs$, and we define $\IndMod{r,s}$ to be the quotient of $\VerMod{r,s}$ by the submodule generated by this singular vector.  $\IndMod{r,s}$ is an in general reducible \hwm{}, but for special values of $r$ and $s$ (when $r=1,2$ and $3$ divides $s$, or when $2$ divides $r$ and $s=1,2,3$) it coincides with the corresponding irreducible module.  Irreducible \hwms{} will be denoted by $\IrrMod{r,s}$.  The \hwss{} of the \hwms{} $\VerMod{r,s}$, $\IndMod{r,s}$ and $\IrrMod{r,s}$ will all be denoted by $\ket{\phi_{r,s}}$ (and we trust that this will not cause any confusion).

\begin{table}
\begin{center}
\setlength{\extrarowheight}{4pt}
\begin{tabular}{|C|C|C|C|C|C|C|C|C|C|C}
\hline
0 & 0 & \tfrac{1}{3} & 1 & 2 & \tfrac{10}{3} & 5 & 7 & \tfrac{28}{3} & 12 & \cdots \\[1mm]
\hline
\tfrac{5}{8} & \tfrac{1}{8} & \tfrac{-1}{24} & \tfrac{1}{8} & \tfrac{5}{8} & \tfrac{35}{24} & \tfrac{21}{8} & \tfrac{33}{8} & \tfrac{143}{24} & \tfrac{65}{8} & \cdots \\[1mm]
\hline
2 & 1 & \tfrac{1}{3} & 0 & 0 & \tfrac{1}{3} & 1 & 2 & \tfrac{10}{3} & 5 & \cdots \\[1mm]
\hline
\vdots & \vdots & \vdots & \vdots & \vdots & \vdots & \vdots & \vdots & \vdots & \vdots & \ddots
\end{tabular}
\vspace{3mm}
\caption{A part of the extended Kac table for $c=0$, displaying the conformal dimensions $h_{r,s}$ (\eqnref{eqnConfDim}) of the primary fields $\phi_{r,s}$.  The rows of the table are labelled by $r = 1, 2, 3, \ldots$ and the columns by $s = 1, 2, 3, \ldots$.} \label{tabExtKacc=0}
\end{center}
\end{table}

We mention that whilst this extended Kac table has been taken as fundamental throughout much of the literature, our first observation (see \secref{secWatts}) is that a \cft{} description of Watts' crossing probability requires the introduction of indecomposable modules which have no natural interpretation in terms of this table.  Instead, we introduce a ``shifted'' extended Kac table (\tabref{tabShExtKacc=0}) in \secref{secWatts} in which the modules required by both Cardy's and Watts' formulae can be accommodated.  This failure of the usual extended Kac table is not surprising in hindsight, and it can be tracked back to the fact that Verma modules of type III$_-$ (see \cite[Thm.~2.2]{FeiVer84} for this and notation) require \emph{two} distinct parallel lines of integral points to describe the complete set of singular vectors.

\section{Cardy's Crossing Probability and LCFT} \label{secLM23}

In this section we shall review the (chiral) structures derived in \cite{RidPer07,RidLog07} from the existence of the boundary condition changing operator $\phi_{1,2}$, required for Cardy's computation of the horizontal crossing probability in critical percolation \cite{CarCri92}.  This primary field was proven to generate a reducible but indecomposable module, which we may identify with $\IndMod{1,2}$.  The structure of this module completely encodes the requirements of Cardy's derivation.

\begin{figure}
\begin{center}
\includegraphics[width=6cm]{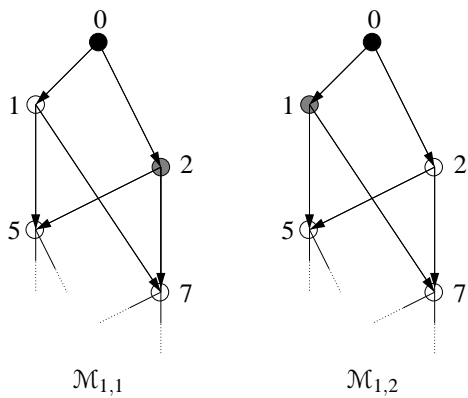}
\caption{A schematic picture of the indecomposable modules $\IndMod{1,1}$ and $\IndMod{1,2}$.  The black circles represent the highest weight states, grey denotes a singular vector that does \emph{not} identically vanish, and white denotes the identically vanishing singular vectors.  These states are labelled by their conformal dimension.} \label{figM11M12}
\end{center}
\end{figure}

The spectrum of the percolation \cft{} must then contain, in the boundary sector, all modules generated from $\IndMod{1,2}$ by fusion.  Using the algebraic fusion algorithm of Nahm and Gaberdiel-Kausch \cite{GabFus94,NahQua94,GabInd96}, this was investigated explicitly\footnote{One cannot use the more standard methods of computing fusion rules (using $3$-point correlation functions for example), because the module $\IndMod{1,2}$ contains non-trivial null states which are orthogonal to the entire module.  The matrices of $2$-point functions at each grade are therefore degenerate (singular), hence we require a method of computing fusion which does not refer to correlation functions.}.  As one might expect, fusing $\IndMod{1,2}$ with itself generates two modules:
\begin{equation} \label{eqnFRM12M12}
\IndMod{1,2} \fuse \IndMod{1,2} = \IndMod{1,1} \oplus \IndMod{1,3}.
\end{equation}
(Here we denote the fusion operation by $\fuse$ to distinguish it from the direct product $\times$.)  The module $\IndMod{1,1}$ is generated by a \hws{} of dimension $0$ which is annihilated by $L_{-1}$.  This is then the vacuum $\ket{0}$, and in fact one finds that
\begin{equation}
\IndMod{1,1} \fuse \IndMod{1,1} = \IndMod{1,1} \qquad \text{and} \qquad \IndMod{1,1} \fuse \IndMod{1,2} = \IndMod{1,2}.
\end{equation}
The vacuum module $\IndMod{1,1}$ therefore serves as the fusion identity (on the fusion subring generated by $\IndMod{1,2}$).  We illustrate the singular vector structures of the modules $\IndMod{1,1}$ and $\IndMod{1,2}$ in \figref{figM11M12}.  This structure is what distinguishes the two dimension $0$ \hwms{}.  Note that $\IndMod{1,1}$ is also reducible but indecomposable.

\psfrag{0}[][]{$0$}
\psfrag{1}[][]{$1$}
\psfrag{2}[][]{$2$}
\psfrag{5}[][]{$5$}
\psfrag{7}[][]{$7$}
\psfrag{M11}[][]{$\IndMod{1,1}$}
\psfrag{M12}[][]{$\IndMod{1,2}$}

We can generate further modules by fusing $\IndMod{1,2}$ with the module $\IndMod{1,3}$ generated by (\ref{eqnFRM12M12}).  Repeating, we deduce that the modules $\IndMod{1,3k} = \IrrMod{1,3k}$ ($k \in \ZZ_+$) appear, as do new modules which we shall denote by $\LogMod{1,3k+1}$ and $\LogMod{1,3k+2}$ ($k \in \ZZ_+$).  The fusion ring is associative (as expected) and the general fusion rules may be obtained from those with the generator $\IndMod{1,2}$:
\begin{subequations} \label{eqnLM23Fusion}
\begin{align}
\IndMod{1,2} \fuse \IndMod{1,3k} &= \LogMod{1,3k+1} & & \\
\IndMod{1,2} \fuse \LogMod{1,3k+1} &= 2 \IndMod{1,3k} \oplus \LogMod{1,3k+2} & &\text{($k \in \ZZ_+$).} \\
\IndMod{1,2} \fuse \LogMod{1,3k+2} &= \IndMod{1,3 \brac{k-1}} \oplus \LogMod{1,3k+1} \oplus \IndMod{1,3 \brac{k+1}} & &
\end{align}
\end{subequations}
(We define $\IndMod{1,0}$ to be the trivial zero-dimensional module.)  These new modules $\LogMod{1,s}$ are examples of rank-$2$ staggered modules \cite{RohRed96}, and it is these which give rise to logarithms in the correlation functions of the theory.  We mention that the modules generated by repeatedly fusing $\IndMod{1,2}$ can be naturally associated with the first row of the $c=0$ extended Kac table (\tabref{tabExtKacc=0}).  We will sometimes refer to them as \emph{first-row} modules in what follows.

The structure of the staggered modules $\LogMod{1,s}$ ($s = 3k + \ell$, $k \in \ZZ_+$, $\ell = 1 , 2$) can be characterised as follows.  Their maximal \hwsm{} is isomorphic to $\IndMod{1,s'}$ (where $s' = 3k - \ell$).  Quotienting the staggered module by this submodule gives another \hwm{}, this time isomorphic to $\IndMod{1,s}$.  (In \cite{RohRed96}, the maximal \hwsm{} and the corresponding quotient \hwm{} are referred to as the ``lower'' and ``upper'' modules, respectively.)  This is summarised mathematically by the short exact sequence
\begin{equation}
\dses{\IndMod{1,3k - \ell}}{\LogMod{1,3k + \ell}}{\IndMod{1,3k + \ell}} \qquad \text{($k \in \ZZ_+$, $\ell = 1 , 2$).}
\end{equation}

\begin{figure}
\begin{center}
\includegraphics[width=12cm]{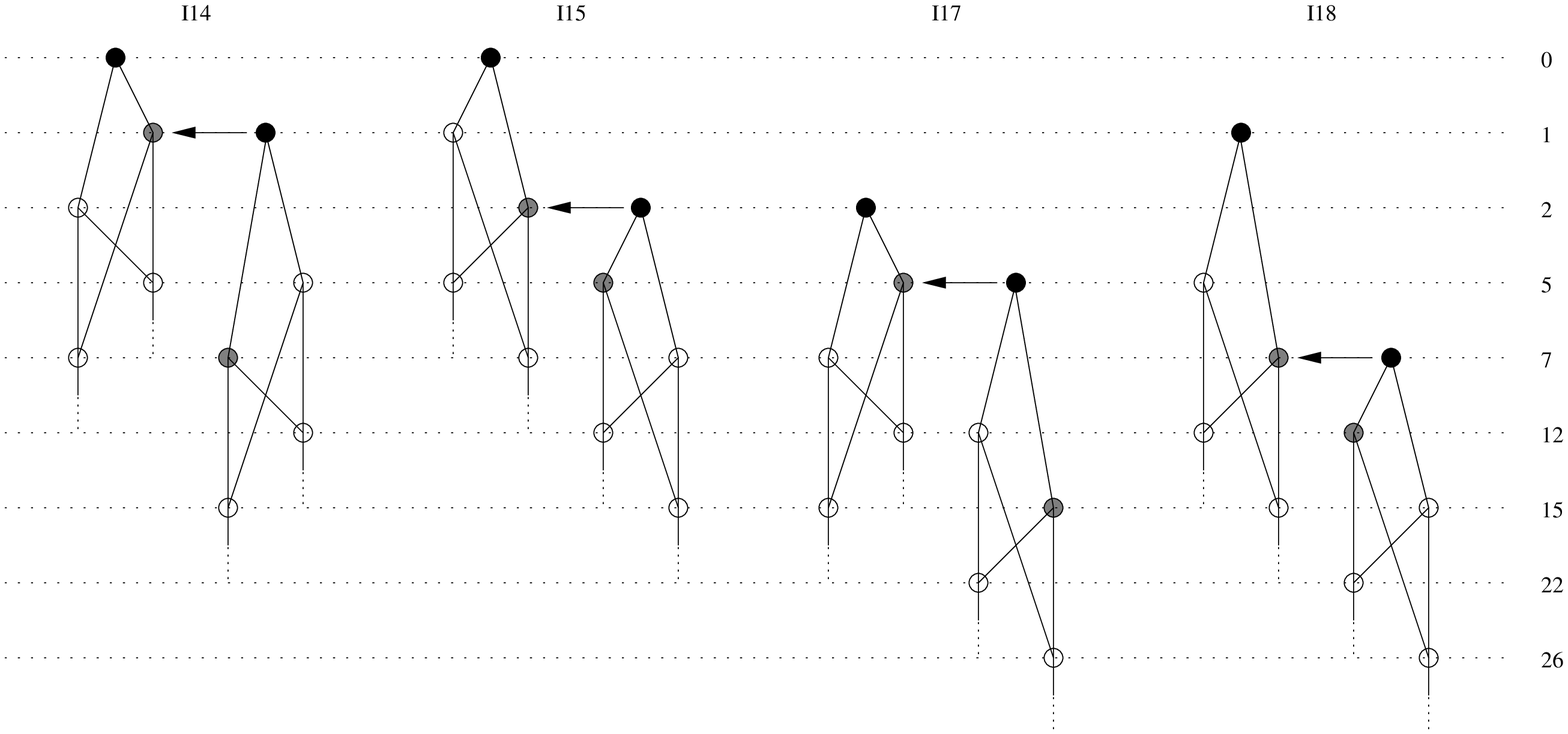}
\caption{The structure of the staggered modules $\LogMod{1,s}$ with $s=4,5,7,8$.  As before, the black circles represent the highest weight states, grey denotes a singular vector that does not identically vanish, and white denotes the identically vanishing singular vectors.  For the circles comprising the right half of each diagram, this is understood to mean that these states are singular in the quotient $\IndMod{1,s}$ (not in the staggered module itself).  The conformal dimension is indicated on the right.  The arrows denote the logarithmic coupling between the state $\ket{\lambda_{1,s}}$ (on the right) and $\ket{\chi_{1,s'}}$ (on the left).} \label{figI1s}
\end{center}
\end{figure}

The submodule $\IndMod{1,s'}$ is itself reducible but indecomposable and contains a unique non-vanishing singular descendant of $\ket{\phi_{1,s'}}$ which will be denoted by $\ket{\chi_{1,s'}}$.  By the usual abuse of notation, we will also denote the corresponding elements of $\LogMod{1,s}$ by $\ket{\phi_{1,s'}}$ and $\ket{\chi_{1,s'}}$.  The quotient module $\IndMod{1,s}$ is likewise reducible, but we cannot identify its \hws{} $\ket{\phi_{1,s}}$ with any element of $\LogMod{1,s}$.  Indeed, its preimage in $\LogMod{1,s}$ is only defined up to the subspace of $\IndMod{1,s'}$ with conformal dimension $h_{1,s}$.  Choosing a representative $\ket{\lambda_{1,s}}$ of this preimage gives a state which is \emph{not} an eigenvector of $L_0$ as $L_0 \ket{\lambda_{1,s}} = h_{1,s} \ket{\lambda_{1,s}}$ need only hold modulo $\IndMod{1,s'}$.  It is a simple exercise to demonstrate \cite{RohRed96} that $\brac{L_0 - h_{1,s} \id} \ket{\lambda_{1,s}}$ must be a \hws{}, so it follows that we may normalise our representative $\ket{\lambda_{1,s}}$ so that
\begin{equation} \label{eqnDefLambda}
L_0 \ket{\lambda_{1,s}} = h_{1,s} \ket{\lambda_{1,s}} + \ket{\chi_{1,s'}},
\end{equation}
at least once a normalisation for $\ket{\chi_{1,s'}}$ has been decided upon.  The states $\ket{\chi_{1,s'}}$ and $\ket{\lambda_{1,s}}$ then constitute a non-diagonalisable Jordan cell\footnote{Normally, $L_0$ is guaranteed to be diagonalisable by virtue of it being self-adjoint.  Whilst $L_0$ is still self-adjoint in logarithmic theories, the inner product on the states spanning this Jordan cell is \emph{indefinite} (neither positive nor negative-definite).  This lack of unitarity --- our state space is not a (pre-)Hilbert space --- resolves the seeming contradiction \cite{BogInd74}.} (in normal form) for $L_0$.  The structures of the first few staggered modules are illustrated in \figref{figI1s}.

\psfrag{0}[][]{$\scriptstyle 0$}
\psfrag{1}[][]{$\scriptstyle 1$}
\psfrag{2}[][]{$\scriptstyle 2$}
\psfrag{5}[][]{$\scriptstyle 5$}
\psfrag{7}[][]{$\scriptstyle 7$}
\psfrag{12}[][]{$\scriptstyle 12$}
\psfrag{15}[][]{$\scriptstyle 15$}
\psfrag{22}[][]{$\scriptstyle 22$}
\psfrag{26}[][]{$\scriptstyle 26$}
\psfrag{I14}[][]{$\LogMod{1,4}$}
\psfrag{I15}[][]{$\LogMod{1,5}$}
\psfrag{I17}[][]{$\LogMod{1,7}$}
\psfrag{I18}[][]{$\LogMod{1,8}$}

As previously mentioned, it is this staggered module structure which signals a \lcft{}.  Indeed, solving the three \pdes{} induced by the conformal invariance of the vacuum shows that the correlation function $\corrfn{\func{\lambda_{1,s}}{z} \func{\lambda_{1,s}}{w}}$ has the form
\begin{equation} \label{eqnCFLamLam}
\corrfn{\func{\lambda_{1,s}}{z} \func{\lambda_{1,s}}{w}} = \frac{C_1 + C_2 \log \brac{z-w}}{\brac{z-w}^{2 h_{1,s}}}
\end{equation}
(in fact, one only needs to solve two of these equations --- the third is then satisfied automatically).  Here, $C_1$ and $C_2$ are constants.  The latter is computable --- it is related to the logarithmic coupling $\beta_{1,s}$ defined below in \eqnref{eqnDefBeta} (in a manner depending upon $s$) --- whereas the former is not \cite{RidPer07}.

We point out that the ``logarithmic partner state'' $\ket{\lambda_{1,s}}$ need not be annihilated by the $L_n$ with $n > 0$, though the result must belong to the submodule $\IndMod{1,s'}$.  Determining the action of the positive Virasoro modes on $\ket{\lambda_{1,s}}$ then completely fixes the structure of the staggered module $\LogMod{1,s}$ as a $\alg{Vir}$-module.  However, this determination is subject to the following subtlety:  The action of these positive modes on $\ket{\lambda_{1,s}}$ is in general not well-defined.  To see this, recall that \eqnref{eqnDefLambda} only defines $\ket{\lambda_{1,s}}$ up to states in $\IndMod{1,s'}$ of dimension $h_{1,s}$.  For $s = 4,5$, the only such state is $\ket{\chi_{1,s'}}$ (and its multiples) which is annihilated by the positive Virasoro modes.  For $s \geqslant 7$ however, there are other states which are not annihilated by these modes, so for $s \geqslant 7$, the action of the positive Virasoro modes on $\ket{\lambda_{1,s}}$ is not well-defined.

Following \cite{RidLog07}, we will refer to the redefinitions
\begin{equation} \label{eqnGTs}
\ket{\lambda_{1,s}} \longmapsto \ket{\lambda_{1,s}} + \ket{\psi}, \qquad L_0 \ket{\psi} = h_{1,s} \ket{\psi},
\end{equation}
as \emph{gauge transformations}.  There is one obvious gauge-invariant quantity that we must compute, the so-called \emph{logarithmic coupling}.  It is most simply defined as
\begin{equation} \label{eqnDefBeta}
\beta_{1,s} = \braket{\chi_{1,s'}}{\lambda_{1,s}},
\end{equation}
and so appears as the coefficient of $\ket{\phi_{1,s'}}$ upon acting on $\ket{\lambda_{1,s}}$ with an appropriate sum of strings of positive Virasoro modes.  (It follows from this observation that the staggered modules $\LogMod{1,s}$ are cyclic, each being generated as a $\alg{Vir}$-module by the state $\ket{\lambda_{1,s}}$.)

This logarithmic coupling may be computed directly from the Nahm-Gaberdiel-Kausch fusion algorithm.  Alternatively \cite{RidLog07}, it may be obtained as a by-product of computing the vanishing singular vector associated with $\ket{\lambda_{1,s}}$.  The corresponding singular vector in $\IndMod{1,s}$ vanishes, so the same combination of Virasoro modes acting on $\ket{\lambda_{1,s}}$ must give an element of the submodule $\IndMod{1,s'}$.  Determining this latter element determines the logarithmic coupling.  In any case, either way quickly becomes tedious, and so far only the logarithmic couplings with $s \leqslant 8$ have been computed explicitly (see \cite{RidLog07} for some conjectured general formulae however):
\begin{equation} \label{eqnOldBetas}
\beta_{1,4} = \frac{-1}{2}, \qquad \beta_{1,5} = \frac{-5}{8}, \qquad \beta_{1,7} = \frac{-35}{3}, \qquad \text{and} \quad \beta_{1,8} = \frac{-13475}{216}.
\end{equation}
These values assume the normalisation for $\ket{\chi_{1,s'}}$ used in \cite{RidPer07,RidLog07} ($\ket{\chi_{1,s'}}$ is expressed as a descendant of $\ket{\phi_{1,s'}}$ in which all terms are Poincar\'{e}-Birkhoff-Witt ordered and such that the coefficient of the term with only one Virasoro mode is unity).  Knowledge of these couplings $\beta_{1,s}$ is necessary in order to compute within the staggered module $\LogMod{1,s}$ and thereby compute correlation functions of the corresponding fields.

We want to emphasise an obvious but important feature of the staggered modules $\LogMod{1,s}$.  Physics mandated the introduction of the indecomposable $\IndMod{1,2}$, but this module is quite unsatisfactory from a field-theoretic point of view.  The problem is the presence of non-vanishing null states, orthogonal to every state in the module.  Correlation functions involving the corresponding fields should therefore vanish identically, so what prevents us from setting these null states to zero?  The answer is that $\IndMod{1,2}$ is actually realised as a submodule of $\LogMod{1,4}$.  In the latter, the non-vanishing null states are paired with logarithmic partners, and each pair is not orthogonal.  There are therefore correlation functions involving the null fields which do not vanish identically.  More abstractly, we see that whereas the module $\IndMod{1,2}$ is not isomorphic to its restricted dual (see \cite{MooLie95} for a definition), the staggered module $\LogMod{1,4}$ is.  Note that this self-duality property of rank-$2$ staggered modules is also shared by the irreducible modules that compose a traditional \cft{}.

The correlation functions of a chiral \lcft{} themselves lead to fundamental issues relating to the theory's consistency.  The simplest example illustrating this, originally considered\footnote{There the consideration was restricted to the level of \opes{}, postulated in order to avoid the $c \rightarrow 0$ catastrophe.} in \cite{GurCon04}, pertains to $\LogMod{1,5}$.  Here, the null field $\func{\chi_{1,1}}{z}$ is the energy-momentum tensor $\func{T}{z}$.  Our three invariant-vacuum \pdes{} give
\begin{equation}
\corrfn{\func{\chi_{1,1}}{z} \func{\lambda_{1,5}}{w}} = \frac{\beta_{1,5}}{\brac{z-w}^4} \qquad \text{and} \qquad \corrfn{\func{\lambda_{1,5}}{z} \func{\lambda_{1,5}}{w}} = \frac{C - 2 \beta_{1,5} \log \brac{z-w}}{\brac{z-w}^4},
\end{equation}
where $C$ is an unknowable (non-gauge-invariant) constant (see \eqnref{eqnCFLamLam}).  Suppose now that we wanted to augment our theory by a staggered module $\LogMod{}$ of a form similar (but different to) that of $\LogMod{1,5}$.  More precisely, suppose that $\LogMod{}$ also has $\IndMod{1,1}$ as its maximal \hwsm{}, and that the quotient $\LogMod{} / \IndMod{1,1}$ is highest weight of dimension $h_{1,5} = 2$.  Consistency is now brought into question through the consideration of the correlation function $\corrfn{\func{\lambda}{z} \func{\lambda_{1,5}}{w}}$, where $\func{\lambda}{z}$ denotes the (logarithmic) Jordan partner field to $\func{\chi_{1,1}}{z} = \func{T}{z}$ in $\LogMod{}$.  It transpires that two of the invariant-vacuum \pdes{} give the solution
\begin{equation}
\corrfn{\func{\lambda}{z} \func{\lambda_{1,5}}{w}} = \frac{C' - \brac{\beta + \beta_{1,5}} \log \brac{z-w}}{\brac{z-w}^4},
\end{equation}
where $\beta$ is the logarithmic coupling of $\LogMod{}$, but this does \emph{not} satisfy the third \pde{} unless $\beta = \beta_{1,5}$.

The suppositions discussed above are not purely academic.  Such a situation occurs if we try to augment our theory by the module $\IndMod{2,1}$ \cite{RidPer07}.  This is a natural assumption in many respects --- fields of dimension $h_{2,1} = \tfrac{5}{8}$ have a long history of being associated with percolation \cite{DotCon84,ChiInt92}.  But, fusing this module with itself leads to a staggered module $\LogMod{} = \LogMod{3,1}$ of precisely the form discussed above.  The logarithmic coupling turns out to be $\beta_{3,1} = \tfrac{5}{6} \neq \tfrac{-5}{8} = \beta_{1,5}$, and so the conformal invariance of the vacuum makes this augmentation \emph{inconsistent}.  Similar arguments can be made for other augmentations by modules of the form $\IndMod{r,s}$ with $r,s \in \ZZ_+$ and $r>1$ \cite{RidLog07}.

One conclusion is then the following:  Cardy's derivation of the horizontal crossing probability forces the \cft{} of critical percolation to contain, in the boundary sector, the chiral \lcft{} generated by the indecomposable module $\IndMod{1,2}$, and moreover, this \emph{chiral} theory cannot be augmented by the modules $\IndMod{r,s}$ with $r>1$.  However, it is necessary to keep in mind that we have only proven that the module $\LogMod{}$ cannot coexist with $\LogMod{1,5}$ in a chiral \lcft{}.  As we shall see (\secref{secDisc}), there are loopholes by which the hypotheses of this theorem can be avoided, but physical relevance maintained.  Nevertheless, it is obvious that any serious proposal for a physical \lcft{} should detail whether consistency issues arise, and how they are resolved if they do.  With this in mind, we now turn to the other crossing probabilities that should also be accommodated within the percolation \cft{}.  In particular, we will consider the probability for simultaneous horizontal and vertical crossings as determined by Watts \cite{WatCro96}.

\section{Watts' Crossing Probability and LCFT} \label{secWatts}

Recall that to derive his horizontal crossing probability, Cardy noted \cite{CarCri92} that it could be expressed as a linear combination of $4$-point functions of the boundary condition changing operators associated with the primary $\phi_{1,2}$.  This identification of the boundary condition changing operators was suggested by an extrapolation of the corresponding identifications in the Ising model and $3$-state Potts model.  Specifically, percolation can be viewed as the $q \rightarrow 1$ limit of the $q$-state Potts model (the Ising model is $q=2$).  Once this identification has been made, the descendant singular vector of $\ket{\phi_{1,2}}$ at grade $2$ induces a second-order \ode{} for the crossing probability.  This is easily solved (given the obvious boundary conditions).

Watts' derivation \cite{WatCro96} of the probability of having simultaneous horizontal and vertical crossings in a rectangle of given shape starts from the assumption that this probability can also be expressed as a linear combination of $4$-point functions of some boundary condition changing operators.  Conformal invariance of this probability requires that the associated primary field have vanishing conformal dimension.  Unfortunately, Watts was not able to propose a candidate for this field in percolation, essentially because the fields corresponding to his boundary condition changing operators were not known in the $q$-state Potts models (with $q=2$ and $3$).  Instead, he was able to derive certain properties that this crossing probability must satisfy.  The solutions to Cardy's second-order \ode{} do not satisfy these properties, but Watts found that there is unique solution to the \emph{fifth}-order \ode{} (induced by the grade $5$ descendant singular vector of $\ket{\phi_{1,2}}$) which does.  Satisfyingly, this solution beautifully interpolates the results of the corresponding numerical simulations \cite{LanUni92} (and has been subsequently proven via \sle{} \cite{DubExc06}).

In hindsight, this success suggests an obvious proposal for the primary field corresponding to Watts' boundary condition changing operators.  In the formalism of \secref{secLM23}, Watts' primary field, $\phi$ say, cannot be identified with $\phi_{1,2} \in \IndMod{1,2}$ (or $\phi_{1,1} \in \IndMod{1,1}$ obviously).  Instead, it seems clear that $\phi$ must belong to an indecomposable dimension $0$ module in which both the singular vectors at grades $1$ and $2$ are non-vanishing, but that at grade $5$ vanishes.  Such a module has not yet appeared in any theory of critical percolation (to our knowledge).  In particular, this means that we must further augment the logarithmic boundary theory of \cite{RidPer07} by this module.  As we have noted, na\"{\i}ve augmentations of this theory frequently lead to inconsistencies, so we will have to carefully analyse the representations induced by this physical augmentation to make sure that mathematical consistency is preserved.

Before continuing with the representation theory, let us gather some further evidence for our proposal for Watts' primary field.  In \cite{SimPer07}, there is an alternative derivation of both Cardy's and Watts' crossing probabilities (among others).  There, the authors consider crossing probability \emph{densities} for which the relevant boundary condition changing operator is argued to correspond to a dimension $1$ primary which the authors denote by $\psi_3$.  There is some discussion of the relation of this field to $\partial \phi_{1,2}$, the derivative of Cardy's field, which is also primary of dimension $1$.  This discussion stems from the fact that recovering the physical crossing probabilities from the corresponding densities requires integrating over the arguments of the $\psi_3$ fields.  However, the two dimension $1$ primaries cannot coincide if $\psi_3$ is to be relevant for Watts' crossing formula.  The interpretation is that $\psi_3$ is to be strictly identified with $\partial \phi_{1,2}$ only when considering Cardy's crossing density.

What is important for our purposes is that the state $\ket{\psi_3}$ has a descendant singular vector at grade $4$ (hence of dimension $5$).  Of course, so does $\ket{\partial \phi_{1,2}} = L_{-1} \ket{\phi_{1,2}}$, but the essential point is that $\ket{\phi_{1,2}}$ has a \emph{vanishing} descendant at dimension $2$, so the singular descendants of $\ket{\partial \phi_{1,2}}$ at dimensions $5$ and $7$ must both vanish (see \figref{figM11M12}).  $\ket{\psi_3}$ is thereby distinguished, representation-theoretically, from $\ket{\partial \phi_{1,2}}$ if its grade $6$ (dimension $7$) singular descendant does not vanish.

\begin{figure}
\begin{center}
\includegraphics[width=2.5cm]{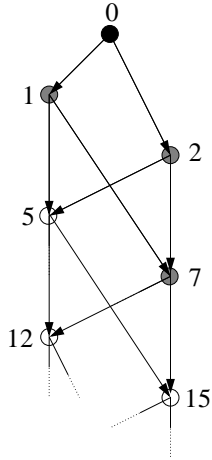}
\caption{A schematic picture of the module generated by Watts' primary field $\phi$.  Again, the black circles represent the highest weight state, grey denotes a non-vanishing singular vector, and white denotes a vanishing one.  These states are labelled by their conformal dimension.} \label{figWattsMod}
\end{center}
\end{figure}

This would seem to settle the issue.  For computing Watts' crossing density, we identify $\psi_3$ with our dimension $1$ primary field $\phi_{1,4}$.  Recall that the corresponding state $\ket{\phi_{1,4}}$ generates the \hwm{} $\IndMod{1,4}$ which is realised in percolation as a submodule of the staggered module $\LogMod{1,8}$ (\figref{figI1s}).  However, this is somewhat unsatisfactory as it gives an interpretation for the $4$-point functions corresponding to Watts' crossing density, but not for those corresponding to his crossing probability.  The latter, as noted above, is obtained by integrating the coordinate of the field $\psi_3$.  We therefore propose that the correct interpretation of the derivation of \cite{SimPer07} is that $\psi_3$ must be identified with a dimension $1$ primary field $\partial \phi$.  $\phi$ must then be a well-defined dimension $0$ field with non-vanishing singular descendants at grades $1$ and $2$, and a vanishing singular descendant at grade $5$.  This is exactly the structure which we have proposed based on Watts' original derivation.  Moreover, we also now know that the singular descendant at grade $7$ should not vanish, so as to distinguish $\partial \phi_{1,2}$ (Cardy's density) from $\partial \phi$ (Watts' density)\footnote{Clearly $\partial \phi$ will be null (unlike $\phi_{1,4}$).  We consider this further evidence in its favour --- $\partial \phi_{1,2}$ is also null, yet computes Cardy's crossing density admirably.}.  We indicate the structure of the module generated by $\ket{\phi}$ in \figref{figWattsMod}.

\psfrag{0}[][]{$0$}
\psfrag{1}[][]{$1$}
\psfrag{2}[][]{$2$}
\psfrag{5}[][]{$5$}
\psfrag{7}[][]{$7$}
\psfrag{12}[][]{$12$}
\psfrag{15}[][]{$15$}

We mention that it is possible to see in a different manner that the singular descendant of $\ket{\phi}$ at grade $7$ must not vanish.  The module obtained from the dimension $0$ Verma module by setting \emph{both} the singular vectors at grade $5$ and $7$ to zero can be fused with itself using the Nahm-Gaberdiel-Kausch algorithm.  Computing to grade $3$ (for example), we find that the resulting fusion contains among others a copy of the module $\IndMod{1,3}$ and a submodule isomorphic to the irreducible vacuum module $\IrrMod{1,1}$.  This result is self-contradictory --- an irreducible vacuum at $c=0$ precludes the existence of any other states \cite{FeiAnn92,RidPer07}, including those of $\IndMod{1,3}$ for instance.

We therefore consider the module obtained from the dimension $0$ Verma module by setting only the grade $5$ singular vector to zero.  This is the module generated by Watts' primary field $\phi$.  Recall that the modules considered in \secref{secLM23} could all be naturally associated with entries in the $c=0$ extended Kac table, \tabref{tabExtKacc=0}.  This is not the case for this new module.  Indeed, identifying the newcomer with an indecomposable of the form $\IndMod{r,s}$, where $h_{r,s} = 0$ and $rs = 5$, yields four solutions for $r$ and $s$ (as usual), but none of these are integral:
\begin{equation}
\brac{r,s} = \pm \brac{2,\tfrac{5}{2}} \qquad \text{and} \qquad \brac{r,s} = \pm \brac{\tfrac{5}{3},3}.
\end{equation}
It turns out to be useful to identify\footnote{As we shall see, this is the appropriate identification for fusing with the first-row modules of \secref{secLM23}, in particular, for fusing with $\IndMod{1,2}$.  If we planned to consider fusions with first-column modules, $\IndMod{2,1}$ for instance, then the identification as $\IndMod{5/3,3}$ would be more useful.} this new module as $\IndMod{2,5/2}$.  We will therefore identify Watts' primary field as $\phi_{2,5/2}$.

In fact, it is convenient to define at this point a \emph{shifted} extended Kac table in which the entries with $r$ even have $s$ half-integral.  We present a part of this shifted extended Kac table in \tabref{tabShExtKacc=0}.  Note that there is a slight redundancy inherent in this table, due to the isomorphisms
\begin{equation} \label{eqnShKacIsos}
\IndMod{2a-1 , 3b} \cong \IndMod{2b , 3a - 3/2}, \qquad a,b \in \ZZ_+.
\end{equation}
This generalises the redundancy in the standard extended Kac table, \tabref{tabExtKacc=0}.  Indeed, we always have $\IndMod{r,s} \cong \IndMod{2s/3,3r/2}$ when this module is well-defined, regardless of the integrality of the indices $r$ and $s$ (and this generalises in the obvious way to other extended Kac tables).

\begin{table}
\begin{center}
\setlength{\extrarowheight}{4pt}
\begin{tabular}{*{21}{C}}
\hline
\multicolumn{2}{|C|}{0} & \multicolumn{2}{|C|}{0} & \multicolumn{2}{|C|}{\tfrac{1}{3}} & \multicolumn{2}{|C|}{1} & \multicolumn{2}{|C|}{2} & \multicolumn{2}{|C|}{\tfrac{10}{3}} & \multicolumn{2}{|C|}{5} & \multicolumn{2}{|C|}{7} & \multicolumn{2}{|C|}{\tfrac{28}{3}} & \multicolumn{2}{|C|}{12} & \cdots \\[1mm]
\hline
 & \multicolumn{2}{|C|}{\tfrac{1}{3}} & \multicolumn{2}{|C|}{0} & \multicolumn{2}{|C|}{0} & \multicolumn{2}{|C|}{\tfrac{1}{3}} & \multicolumn{2}{|C|}{1} & \multicolumn{2}{|C|}{2} & \multicolumn{2}{|C|}{\tfrac{10}{3}} & \multicolumn{2}{|C|}{5} & \multicolumn{2}{|C|}{7} & \multicolumn{2}{|C}{\cdots} \\[1mm]
\hline
\multicolumn{2}{|C|}{2} & \multicolumn{2}{|C|}{1} & \multicolumn{2}{|C|}{\tfrac{1}{3}} & \multicolumn{2}{|C|}{0} & \multicolumn{2}{|C|}{0} & \multicolumn{2}{|C|}{\tfrac{1}{3}} & \multicolumn{2}{|C|}{1} & \multicolumn{2}{|C|}{2} & \multicolumn{2}{|C|}{\tfrac{10}{3}} & \multicolumn{2}{|C|}{5} & \cdots \\[1mm]
\hline
 & \multicolumn{2}{|C|}{\vdots} & \multicolumn{2}{|C|}{\vdots} & \multicolumn{2}{|C|}{\vdots} & \multicolumn{2}{|C|}{\vdots} & \multicolumn{2}{|C|}{\vdots} & \multicolumn{2}{|C|}{\vdots} & \multicolumn{2}{|C|}{\vdots} & \multicolumn{2}{|C|}{\vdots} & \multicolumn{2}{|C|}{\vdots} & \multicolumn{2}{|C}{\ddots}
\end{tabular}
\vspace{3mm}
\caption{A part of the shifted extended Kac table for $c=0$, displaying the conformal dimensions $h_{r,s}$ (\eqnref{eqnConfDim}) of the primary fields $\phi_{r,s}$.  The rows of the table are labelled by $r = 1, 2, 3, \ldots$ and the columns by $s$.  When $r$ is odd, $s$ takes values $1, 2, 3, \ldots$, but when $r$ is even, $s$ takes values $\tfrac{3}{2}, \tfrac{5}{2}, \tfrac{7}{2}, \ldots$.} \label{tabShExtKacc=0}
\end{center}
\end{table}

We now turn to the fusion rules of $\IndMod{2,5/2}$.  As we are proposing to augment the spectrum required by Cardy's crossing formula by this module, we will focus first upon its fusion with the first-row modules.  In particular, we find that $\IndMod{1,1}$ is again verified to act as the fusion identity.  We can therefore turn to the fusion with $\IndMod{1,2}$.  Computing at grade $0$, we find that the fusion product is generated by two \hwss{} of dimensions $0$ and $\tfrac{1}{3}$.  The first vanishing singular vector is found at grade $3$, identifying $\IndMod{1,3}$ as a direct summand.  The next vanishing singular vector does not occur until grade $7$, and reveals itself to be a descendant of the dimension $0$ \hws{}.  This then identifies the second summand in the fusion decomposition as a new module which we may identify as $\IndMod{2,7/2}$ (note that $h_{2,7/2} = 0$).  We have therefore derived the fusion rule
\begin{equation}
\IndMod{1,2} \fuse \IndMod{2,5/2} = \IndMod{2,3/2} \oplus \IndMod{2,7/2},
\end{equation}
where \eqnref{eqnShKacIsos} has been used to identify $\IndMod{1,3}$ with $\IndMod{2,3/2}$.

\begin{figure}
\begin{center}
\includegraphics[width=8cm]{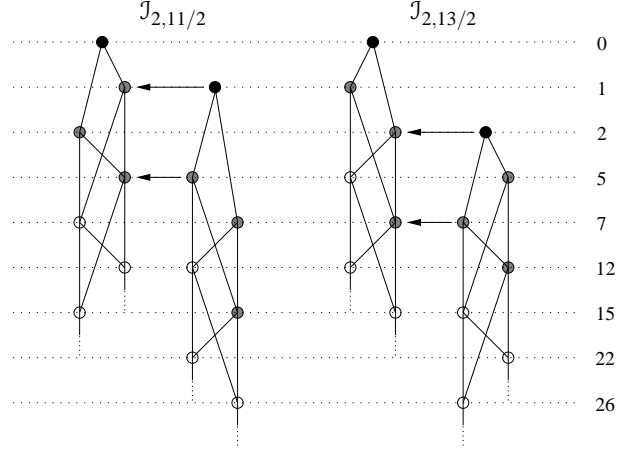}
\caption{The structure of the staggered modules $\LogMod{2,s}$ with $s=11/2$ and $13/2$.  As before, the black circles represent the highest weight states, grey denotes a singular vector that does not identically vanish, and white denotes the identically vanishing singular vectors.  For the circles comprising the right half of each diagram, this is understood as in \figref{figI1s}.  The conformal dimension is indicated on the right and the arrows denote the logarithmic coupling between the singular vectors (left) and their logarithmic partners (right).} \label{figI2s}
\end{center}
\end{figure}

We can continue our exploration of the spectrum by fusing $\IndMod{1,2}$ with the newly generated module $\IndMod{2,7/2}$.  The result is perhaps not unexpected\footnote{Actually, computing with the Nahm-Gaberdiel-Kausch algorithm to grade $9$ (necessary to identify $\IndMod{2,9/2}$) was not feasible with our current implementation.  It is easy to rule out $\IndMod{2,3/2}$ as the dimension $\tfrac{1}{3}$ direct summand of this fusion rule.  The only possibilities are then $\IndMod{2,9/2} , \IndMod{3,6} , \IndMod{4,15/2} , \ldots$, of which the first is overwhelmingly likely.  In fact, no fusion rule considered in this section was computed to a grade greater than $7$, so the identification of the deeper structure of these decompositions remains conjectural.}:
\begin{equation}
\IndMod{1,2} \fuse \IndMod{2,7/2} = \IndMod{2,5/2} \oplus \IndMod{2,9/2},
\end{equation}
where $\IndMod{2,9/2} = \IndMod{3,3}$.  Similarly,
\begin{equation}
\IndMod{1,2} \fuse \IndMod{2,9/2} = \LogMod{2,11/2} \qquad \text{and} \qquad  \IndMod{1,2} \fuse \LogMod{2,11/2} = 2 \IndMod{2,9/2} \oplus \LogMod{2,13/2},
\end{equation}
where $\LogMod{2,11/2}$ and $\LogMod{2,13/2}$ are rank-$2$ staggered modules defined by their respective short exact sequences
\begin{align}
&\dses{\IndMod{2,7/2}}{\LogMod{2,11/2}}{\IndMod{2,11/2}} \\
\text{and} \qquad &\dses{\IndMod{2,5/2}}{\LogMod{2,13/2}}{\IndMod{2,13/2}},
\end{align}
and logarithmic couplings
\begin{equation} \label{eqnNewBetas}
\beta_{2,11/2} = \frac{-1}{2} \qquad \text{and} \qquad \beta_{2,13/2} = \frac{-5}{8}.
\end{equation}
These logarithmic couplings are normalised in the same way as in \secref{secLM23}.  Note that they take the same values as $\beta_{1,4}$ and $\beta_{1,5}$ respectively (\eqnref{eqnOldBetas}).  We illustrate these modules in \figref{figI2s}.  It should be clear now why the \emph{shifted} extended Kac table of \tabref{tabShExtKacc=0} was introduced, and we will find it convenient to refer to the modules $\IndMod{r,s}$ with $r=2$ as \emph{second-row} modules in what follows.

\psfrag{0}[][]{$\scriptstyle 0$}
\psfrag{1}[][]{$\scriptstyle 1$}
\psfrag{2}[][]{$\scriptstyle 2$}
\psfrag{5}[][]{$\scriptstyle 5$}
\psfrag{7}[][]{$\scriptstyle 7$}
\psfrag{12}[][]{$\scriptstyle 12$}
\psfrag{15}[][]{$\scriptstyle 15$}
\psfrag{22}[][]{$\scriptstyle 22$}
\psfrag{26}[][]{$\scriptstyle 26$}
\psfrag{I14}[][]{$\LogMod{2,11/2}$}
\psfrag{I15}[][]{$\LogMod{2,13/2}$}

These fusion rules are reminiscent of those of $\IndMod{1,2}$ with the other first-row modules, given in \eqnref{eqnLM23Fusion}.  It is therefore reasonable to conjecture the following:  Fusing the first-row modules with $\IndMod{2,5/2}$ generates the second-row modules $\IndMod{2,5/2}$, $\IndMod{2,7/2}$, $\IndMod{2,3 \brac{k+1/2}}$, $\LogMod{2,3 \brac{k+1/2} + 1}$ and $\LogMod{2,3 \brac{k+1/2} + 2}$ ($k \in \ZZ_+$).  These new staggered modules are defined by the short exact sequences
\begin{equation}
\dses{\IndMod{2,3 \brac{k+1/2} - \ell}}{\LogMod{2,3 \brac{k+1/2} + \ell}}{\IndMod{2,3 \brac{k+1/2} + \ell}} \qquad \text{($k \in \ZZ_+$, $\ell = 1,2$).}
\end{equation}
The corresponding fusion rules can be computed from those of $\IndMod{1,2}$ and associativity (remembering that $\IndMod{2,3/2} = \IndMod{1,3}$).  We conjecture that these rules take the form
\begin{subequations} \label{eqnAugFusion}
\begin{align}
\IndMod{1,2} \fuse \IndMod{2,3 \brac{k+1/2}} &= \LogMod{2,3 \brac{k+1/2} + 1} & & \\
\IndMod{1,2} \fuse \LogMod{2,3 \brac{k+1/2} + 1} &= 2 \IndMod{2,3 \brac{k+1/2}} \oplus \LogMod{2,3 \brac{k+1/2} + 2} & &\text{($k \in \ZZ_+$)} \\
\IndMod{1,2} \fuse \LogMod{2,3 \brac{k+1/2} + 2} &= \IndMod{2,3 \brac{k-1/2}} \oplus \LogMod{2,3 \brac{k+1/2} + 1} \oplus \IndMod{2,3 \brac{k+3/2}}. & &
\end{align}
\end{subequations}
We note that such fusions of first and second-row modules will not always decompose into second-row modules.  For example, associativity gives
\begin{equation}
\IndMod{1,3} \fuse \IndMod{2,5/2} = \LogMod{1,4} \oplus \IndMod{2,9/2}.
\end{equation}
Na\"{\i}vely, we might have expected that the result would decompose (as vector spaces) into the second-row modules $\IndMod{2,1/2}$, $\IndMod{2,5/2}$ and $\IndMod{2,9/2}$.  But, $\IndMod{2,1/2}$ is not defined\footnote{Such a module would have to have a \hws{} of dimension $h_{2,1/2} = 1$ and a vanishing singular vector of dimension $2$.  There is no such Virasoro module.}.  Instead, the identification of $\IndMod{2,3/2}$ and $\IndMod{1,3}$ causes the decomposition to ``spill over'' into the first row.  $\IndMod{2,1/2}$ and $\IndMod{2,5/2}$ are replaced by $\IndMod{1,2}$ and $\IndMod{1,4}$ respectively, and these together are replaced at the level of modules by $\LogMod{1,4}$.

Note that the new staggered modules we have discovered do not lead to any inconsistencies of the type discussed in \secref{secLM23}.  Indeed, it is not hard to prove that associativity prevents the generation of any staggered module composed of one first-row and one second-row module.  The maximal \hwsms{} of the staggered modules found to date are therefore all distinct.  This consistency check is necessary for our proposed construction of a theory encompassing both Cardy's and Watts' crossing probability formulae, but there remain the fusion rules of the second-row modules with one another to investigate.  

These fusion rules are unsurprisingly more complicated, and we will only be able to obtain partial information about them; we defer the discussion to \secref{secFusion}.  For now, we want to comment on certain deficiencies of the second-row modules we have introduced.  The module $\IndMod{2,5/2}$ that we started from is, like $\IndMod{1,2}$, field-theoretically unsatisfactory as it contains non-vanishing states which are orthogonal to the entire module (the corresponding null fields should therefore be zero in every correlation function).  Unlike the case of $\IndMod{1,2} \subset \LogMod{1,4}$, this is not rectified by embedding it into the rank-$2$ staggered module $\LogMod{2,13/2}$.  For this embedding does not provide any partner for the non-vanishing singular vector at grade $1$ (\figref{figI2s}).  Those at grades $2$ and $7$ are partnered, but only the grade $2$ partner has non-vanishing inner product with the singular vector.  The grade $7$ singular vector of $\IndMod{2,5/2} \subset \LogMod{2,13/2}$ is therefore also problematic.  We mention that this problem is not restricted to $\LogMod{2,13/2}$.  Every second-row staggered module exhibits this deficiency.

We have seen then that the staggered module $\LogMod{2,13/2}$ has non-vanishing states orthogonal to the entire module.  We might therefore expect that the corresponding fields vanish in all correlation functions.  But this cannot be.  If it were true, then there would be no obstacle to setting these fields to zero, hence these troublesome states to zero as well.  However, if we set the grade $1$ (and hence grade $7$) singular vector of $\IndMod{2,5/2}$ to zero, we will recover the module $\IndMod{1,2}$.  In other words, if $\LogMod{2,13/2}$ was the largest indecomposable containing $\IndMod{2,5/2}$, then the field theory would allow us to replace Watts' field $\phi_{2,5/2}$ with Cardy's $\phi_{1,2}$.  The resolution is clear --- there must exist an indecomposable extension of $\IndMod{2,5/2}$ which is larger than $\LogMod{2,13/2}$.  This is in stark contrast to what we would have expected given the analogous behaviour of $\IndMod{1,2}$ and the other first-row modules.  We will return to this realisation in the following sections.

\section{Further Fusions} \label{secFusion}

We consider first the fusion of the module $\IndMod{2,5/2}$ with itself.  Proceeding as always with the algorithm of Nahm and Gaberdiel-Kausch, we find that the decomposition to grade $0$ gives five generating states of dimensions $0$, $0$, $1$, $2$ and $\tfrac{1}{3}$.  What is of immediate interest here is that the two dimension $0$ states form a non-trivial Jordan cell.  This situation, in which the \hws{} is itself part of a Jordan cell, has not been previously observed in the theory we are exploring.  Nevertheless, such couplings are by no means uncommon among general \lcfts{}.

We continue computing the fusion decomposition at deeper grades.  At grade $1$, the \hws{} of dimension $0$ is found to have a vanishing descendant singular vector, hence may be identified as the vacuum $\ket{0}$.  Its logarithmic partner, $\ket{\lambda}$ say, does not.  Satisfyingly, we observe that the dimension $2$ descendant singular vector of the vacuum does not vanish, meaning that the vacuum (sub)module appearing in this decomposition is the first-row module $\IndMod{1,1}$, and not its irreducible (but inconsistent) counterpart.

\begin{figure}
\begin{center}
\includegraphics[width=14cm]{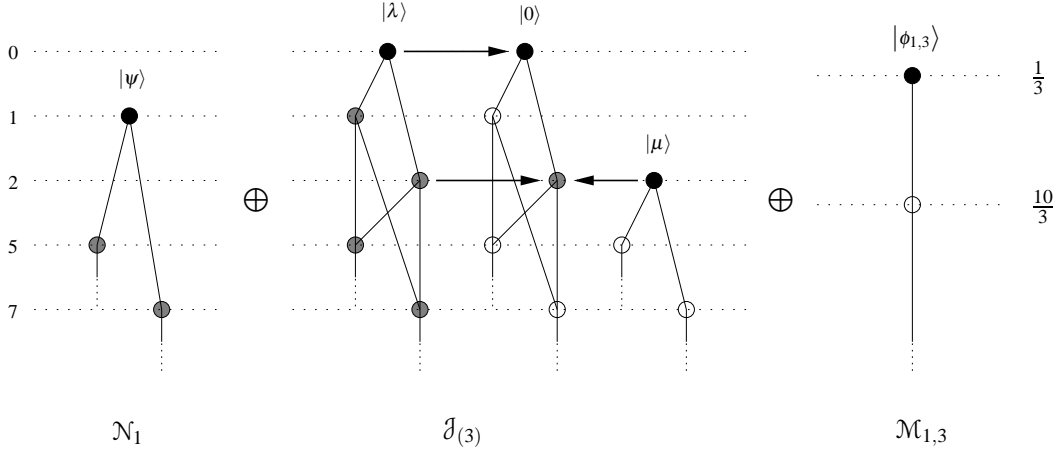}
\caption{The singular vector structure of the modules obtained in the fusion of the module $\IndMod{2,5/2}$ with itself (up to grade $7$).  As usual, the black circles represent the highest weight states, grey denotes singular vectors that do not identically vanish, and white denotes the identically vanishing singular vectors (in the appropriate quotient modules).  Arrows again denote logarithmic coupling.  We mention that the space of logarithmic partner states of dimension $2$ is just $1$, despite the two arrows drawn here.} \label{figWxWDecomp}
\end{center}
\end{figure}

Computing to deeper grades, one uncovers a wealth of vanishing and non-vanishing singular vectors.  In particular, the grade $3$ singular vector descended from the \hws{} of dimension $\tfrac{1}{3}$ is seen to vanish, hence $\IndMod{1,3}$ occurs as a direct summand in the fusion decomposition.  Two more vanishing singular vectors occur at grades $3$ and $5$, both descended from the generator of dimension $2$.  We find no further vanishing singular vectors up to grade $7$ (except for those whose vanishing is forced by what we have already discovered).  The decomposition is therefore as in \figref{figWxWDecomp}.

\psfrag{0}[][]{$\scriptstyle 0$}
\psfrag{1}[][]{$\scriptstyle 1$}
\psfrag{2}[][]{$\scriptstyle 2$}
\psfrag{5}[][]{$\scriptstyle 5$}
\psfrag{7}[][]{$\scriptstyle 7$}
\psfrag{1/3}[][]{$\tfrac{1}{3}$}
\psfrag{10/3}[][]{$\tfrac{10}{3}$}
\psfrag{R3}[][]{$\MegaMod{\brac{3}}$}
\psfrag{M13}[][]{$\IndMod{1,3}$}
\psfrag{N}[][]{$\mathcal{N}_1$}
\psfrag{+}[][]{$\bigoplus$}
\psfrag{vac}[][]{$\scriptstyle \left| 0 \right>$}
\psfrag{lam}[][]{$\scriptstyle \left| \lambda \right>$}
\psfrag{psi}[][]{$\scriptstyle \left| \psi \right>$}
\psfrag{mu}[][]{$\scriptstyle \left| \mu \right>$}
\psfrag{phi}[][]{$\scriptstyle \left| \phi_{1,3} \right>$}

There is a certain delicacy to this analysis however.  Whilst we know that there are generating states of dimensions $1$ and $2$, we cannot identify them uniquely (the same is true for the dimension $0$ state $\ket{\lambda}$, but this is just due to the familiar gauge transformations of \eqnref{eqnGTs}).  In particular, the dimension $1$ generator $\ket{\psi}$ can always be redefined by adding multiples of the other $L_0$-eigenstate of this dimension, $L_{-1} \ket{\lambda}$ (see \figref{figWxWDecomp}).  The situation for the dimension $2$ generator $\ket{\mu}$ is even more complicated as there are three other $L_0$-eigenstates (and one non-eigenstate) of this dimension.  It is easy to check that $\ket{\psi}$ may be \emph{chosen} so that it is annihilated by $L_1$ (and trivially by $L_2$).  This choice makes it a \hws{}.  However, for $\ket{\mu}$, there is no such choice.  We can find a dimension $2$ generating eigenstate which is annihilated by $L_1$, but the best we can do with respect to the $L_2$-action is to have it yield a non-trivial multiple of the vacuum.

One way to understand this is to consider the explicit form of the vanishing grade $3$ singular vector associated with $\ket{\mu}$ (its dimension is thus $5$).  This is found to have components which are strict descendants of this generator as well as (grade $5$) descendants of the vacuum \emph{and} its logarithmic partner $\ket{\lambda}$.  (There are no components descended from $\ket{\psi}$.)  We can choose to eliminate all the components descended from $\ket{\lambda}$ by allowing ourselves to further redefine $\ket{\mu}$ through the addition of multiples of the only state of this dimension which is not an eigenstate of $L_0$.  Once we have done this, $\ket{\mu}$ is then logarithmically coupled to $\ket{\chi_{1,1}} = L_{-2} \ket{0}$ (\figref{figWxWDecomp}).  Normalising it so that $\brac{L_0 - 2 \id} \ket{\mu} = \ket{\chi_{1,1}}$, we can then determine the logarithmic coupling.  It turns out to be $\tfrac{5}{6}$.

This should set off alarm bells.  Recall from \secref{secLM23} that the vacuum module is realised as a submodule of the staggered module $\LogMod{1,5}$.  The logarithmic coupling there was $\beta_{1,5} = \braket{\chi_{1,1}}{\lambda_{1,5}} = \tfrac{-5}{8}$ (\eqnref{eqnOldBetas}).  We therefore have two logarithmic partner states to the same singular vector $\ket{\chi_{1,1}}$ with \emph{different} logarithmic couplings.  This then suggests that we have an inconsistency.  Specifically, we expect that the $2$-point function of $\func{\lambda_{1,5}}{z}$ and $\func{\mu}{w}$ will not satisfy the \pdes{} induced by the conformal invariance of the vacuum.

Let us ignore this problem for the moment, in order to further analyse the situation.  It was noted in \cite{RidLog07} that the logarithmic couplings of these staggered modules were completely determined by the vanishing singular vectors associated with the logarithmic partner states.  Indeed, we chose $\ket{\mu}$ above so that its associated vanishing singular vector at grade $3$ only involved descendants of $\ket{\mu}$ and $\ket{0}$.  The form of this singular vector is the same as that of the module $\LogMod{3,1} = \IndMod{2,1} \fuse \IndMod{2,1}$ (with $\mu$ replaced by $\lambda_{3,1}$), introduced at the end of \secref{secLM23}, which is why we derive $\beta_{3,1} = \tfrac{5}{6}$ for the logarithmic coupling above.

However, the present situation is more complicated in that $\ket{\mu}$ has another associated (vanishing) singular vector at grade $5$ (hence dimension $7$).  If this singular vector only involved descendants of $\ket{\mu}$ and $\ket{0}$, then its form would be identical to that of $\LogMod{1,5}$, and so we would derive $\beta_{1,5} = \tfrac{-5}{8}$ for the logarithmic coupling.  That we do not implies that the form of the grade $5$ singular vector must involve further descendants, and we can check its explicit form to verify that it does indeed involve descendants of $\ket{\lambda}$.

This begs the question:  Can we redefine $\ket{\mu}$ so that the descendants of $\ket{\lambda}$ in the associated grade $5$ singular vector all vanish?  And of course, we can, by adding to $\ket{\mu}$ an appropriate multiple of $L_{-2} \ket{\lambda}$.  For this $\ket{\mu}$, the logarithmic coupling is $\tfrac{-5}{8}$ (and the grade $3$ singular vector has components descended from $\ket{\lambda}$).  Note that this does not resolve the inconsistency induced by having different logarithmic couplings --- these ``other definitions'' of $\ket{\mu}$ are still states of the theory.  Rather, it can be interpreted as expressing that the logarithmic coupling is not gauge-invariant in this more complicated module.  Indeed, redefining $\ket{\mu}$ by adding arbitrary multiples of $L_{-2} \ket{\lambda}$, we can tune the logarithmic coupling to any value we desire.

At the root of this state of affairs is the following observation:  Every state of the form $\brac{L_{-2} + a L_{-1}^2} \ket{\lambda}$ is a (normalised) logarithmic partner to $\ket{\chi_{1,1}} = L_{-2} \ket{0}$.  But,
\begin{equation}
L_2 \brac{L_{-2} + a L_{-1}^2} \ket{\lambda} = \brac{4+6a} \ket{0}.
\end{equation}
This traces the lack of gauge-invariance to the fact that we have logarithmic partner states which are themselves descendants.  Such a situation did not arise when we were only considering first-row staggered modules, but it seems that this is unavoidable in general.

How then can all this inconsistency and lack of gauge-invariance be physical?  The answer lies in the following simple computation:
\begin{equation} \label{eqnNullVac}
\braket{0}{0} = \bracket{0}{L_0}{\lambda} = \bracket{\lambda}{L_0}{0}^* = 0.
\end{equation}
This merely expresses the fact that the eigenstate in a (non-trivial) Jordan cell (for a self-adjoint operator) is necessarily null \cite{BogInd74}.  Its ramifications are, however, huge.  For a null vacuum implies that all the logarithmic couplings actually vanish:
\begin{equation}
L_2 \ket{\mu} \propto \ket{0} \qquad \Rightarrow \qquad \beta = \bracket{0}{L_2}{\mu} = 0 \qquad \text{for all choices of $\ket{\mu}$.}
\end{equation}
In this way, the $2$-point functions of the various logarithmic partner fields become consistent with global conformal invariance of the vacuum.  In fact, these $2$-point functions all vanish identically\footnote{Whilst this observation does solve the consistency issue discussed here, we point out that this ``mass vanishing'' of $2$-point functions means that we are again in the physically unsatisfactory position of having non-zero states orthogonal to the entire module.  Presumably, this means that the module we are considering here is not fully extended, and is actually realised as a submodule of a much larger indecomposable.}.

We have therefore derived the fusion rule
\begin{equation} \label{eqnFusM22.5xM22.5}
\IndMod{2,5/2} \fuse \IndMod{2,5/2} = \mathcal{N}_1 \oplus \MegaMod{\brac{3}} \oplus \IndMod{1,3},
\end{equation}
though we have only been able to derive partial information about the structures of the modules $\mathcal{N}_1$ and $\MegaMod{\brac{3}}$.  In particular, we know that $\mathcal{N}_1$ is a dimension $1$ \hwm{}, but is neither $\IndMod{1,4}$, $\IndMod{3,2}$ nor $\IrrMod{1,4} = \IrrMod{3,2}$.  The latter module $\MegaMod{\brac{3}}$ is a more complicated indecomposable generated by three independent states (\figref{figWxWDecomp}).  We have already identified the vacuum as one of these states, and it generates a submodule of $\MegaMod{\brac{3}}$ isomorphic to $\IndMod{1,1}$.  The partner state to the vacuum $\ket{\lambda}$ heads a module $\mathcal{N}_0$ of unknown character, and the third generator $\ket{\mu}$ gives a copy of $\IrrMod{1,5} = \IrrMod{3,1}$.  More precisely, we can introduce an increasing chain of submodules of $\MegaMod{\brac{3}}$ of the form
\begin{equation} \label{eqnR4CompSeries}
0 = \MegaMod{\brac{0}} \subset \MegaMod{\brac{1}} \subset \MegaMod{\brac{2}} \subset \MegaMod{\brac{3}},
\end{equation}
where the $\MegaMod{\brac{i}}$ are generated by the first $i$ states from the ordered list $\sqbrac{\ket{0}, \ket{\lambda}, \ket{\mu}}$.  It is the quotients $\MegaMod{\brac{i}} / \MegaMod{\brac{i-1}}$, $i=1,2,3$, which we can identify\footnote{Actually, we are assuming here that the vanishing singular vectors associated with $\ket{\lambda}$ (if there are any) are only composed of descendants of $\ket{\lambda}$ and $\ket{0}$.} as
\begin{equation}
\IndMod{1,1}, \qquad \mathcal{N}_0 \qquad \text{and} \qquad \IrrMod{1,5},
\end{equation}
respectively.  The submodule chain (\ref{eqnR4CompSeries}) thus realises a generalised composition series in which irreducibles are replaced by maximal \hwsms{}.

\begin{figure}
\begin{center}
\includegraphics[width=15cm]{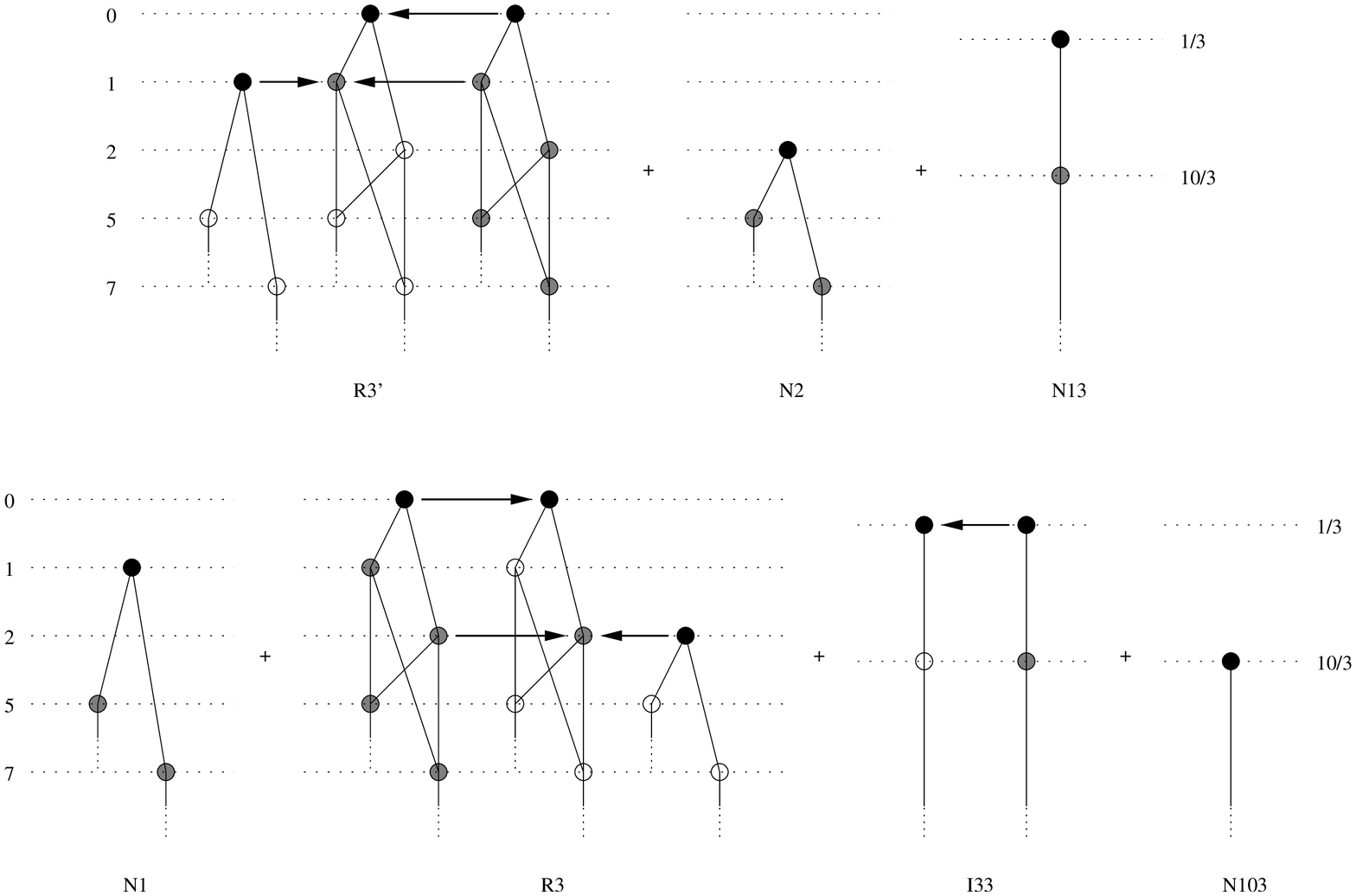}
\caption{The singular vector structure of the modules obtained in the fusion of the modules $\IndMod{2,5/2}$ and $\IndMod{2,7/2}$ (top) and of the module $\IndMod{2,7/2}$ with itself (bottom).  Again, the space of logarithmic partner states of dimensions $1$ and $2$ (respectively) is just $1$, despite the two arrows drawn here.} \label{figMoreDecomps}
\end{center}
\end{figure}

We can similarly investigate the decompositions obtained upon fusing other second-row modules.  For example, we have derived the following rules:
\begin{align}
\IndMod{2,5/2} \fuse \IndMod{2,7/2} &= \mathcal{N}_2 \oplus \MegaMod{\brac{3}}' \oplus \mathcal{N}_{1/3}, \\
\IndMod{2,7/2} \fuse \IndMod{2,7/2} &= \mathcal{N}_1 \oplus \MegaMod{\brac{3}} \oplus \LogMod{1/3} \oplus \mathcal{N}_{10/3}. \label{eqnFusM23.5xM23.5}
\end{align}
The partial characterisations of these modules that we have been able to glean are summarised in \figref{figMoreDecomps}.  In particular, we have found a module $\MegaMod{\brac{3}}'$ similar to $\MegaMod{\brac{3}}$, except that the quotients of its generalised composition series are (in order):
\begin{equation}
\IndMod{1,2}, \qquad \mathcal{N}'_0 \qquad \text{and} \qquad \IrrMod{1,4} = \IrrMod{3,2}.
\end{equation}
$\mathcal{N}'_0$ is another dimension $0$ \hwm{} with no vanishing singular vectors to grade $7$.  We cannot tell if it is isomorphic to $\mathcal{N}_0$ or not.  A plausible guess would be that $\mathcal{N}_0 = \IndMod{3,4}$ and $\mathcal{N}'_0 = \IndMod{3,5}$, but we have no evidence for this.  Similarly, we conjecture that the module we have denoted by $\LogMod{1/3}$ is actually $\LogMod{3,3}$ (that is, the logarithmic singular vector at grade $9$ vanishes).

\psfrag{0}[][]{$\scriptstyle 0$}
\psfrag{1}[][]{$\scriptstyle 1$}
\psfrag{2}[][]{$\scriptstyle 2$}
\psfrag{5}[][]{$\scriptstyle 5$}
\psfrag{7}[][]{$\scriptstyle 7$}
\psfrag{1/3}[][]{$\tfrac{1}{3}$}
\psfrag{10/3}[][]{$\tfrac{10}{3}$}
\psfrag{R3}[][]{$\MegaMod{\brac{3}}$}
\psfrag{R3'}[][]{$\MegaMod{\brac{3}}'$}
\psfrag{N13}[][]{$\mathcal{N}_{1/3}$}
\psfrag{N2}[][]{$\mathcal{N}_2$}
\psfrag{N1}[][]{$\mathcal{N}_1$}
\psfrag{N103}[][]{$\mathcal{N}_{10/3}$}
\psfrag{I33}[][]{$\LogMod{1/3}$}
\psfrag{+}[][]{$\bigoplus$}
\psfrag{+}[][]{$\bigoplus$}
\psfrag{vac}[][]{$\scriptstyle \left| 0 \right>$}
\psfrag{lam}[][]{$\scriptstyle \left| \lambda \right>$}
\psfrag{psi}[][]{$\scriptstyle \left| \psi \right>$}
\psfrag{mu}[][]{$\scriptstyle \left| \mu \right>$}
\psfrag{phi}[][]{$\scriptstyle \left| \phi_{1,3} \right>$}

It is of course possible to use the associativity of the fusion product to try to obtain further results.  This approach can be of some small value.  For example, \eqnref{eqnFusM22.5xM22.5} implies the relations
\begin{align}
\IndMod{1,2} \fuse \brac{\mathcal{N}_1 \oplus \MegaMod{\brac{3}}} &= \Bigl( \IndMod{2,5/2} \fuse \IndMod{2,7/2} \Bigr) \oplus \IndMod{2,9/2}, \\
\text{and} \qquad \IndMod{1,3} \fuse \brac{\mathcal{N}_1 \oplus \MegaMod{\brac{3}}} \oplus \mathcal{N}_1 \oplus \MegaMod{\brac{3}} &= \Bigl( \IndMod{2,7/2} \fuse \IndMod{2,7/2} \Bigr) \oplus \IndMod{1,3} \oplus 2 \LogMod{2,11/2},
\end{align}
the latter of which proves that the modules $\mathcal{N}_1$ and $\MegaMod{\brac{3}}$ appearing in \eqnref{eqnFusM23.5xM23.5} coincide with those of \eqnref{eqnFusM22.5xM22.5}.

\section{Discussion} \label{secDisc}

It is now clear that the \cft{} describing critical percolation is far richer in content than what has been previously suggested.  In particular, we have identified the module generated by the dimension $0$ primary field describing Watts' crossing probability as $\IndMod{2,5/2}$.  Just as with Cardy's probability, this module is reducible but indecomposable, but in contrast, it does not naturally fit into the extended Kac table at $c=0$.  Instead, we have introduced a ``shifted'' extended Kac table which incorporates this module and the other modules known to be present in the theory from the analysis of Cardy's probability (which all appear in the first row).  The point is that fusing the latter first-row modules with $\IndMod{2,5/2}$ fills out the second row of the shifted Kac table, and we have conjectured the general form of these fusion rules.  This suggests that there may in fact exist a sequence of increasingly rarefied crossing probabilities corresponding to the other dimension $0$ modules in the shifted table.  Their interpretation, assuming this speculation has some merit, is an interesting problem, though perhaps not a pressing one.

We have continued our analysis of the percolation theory by fusing the simplest second-row modules with one another.  What we have found is that the resulting indecomposable modules have a far more intricate structure.  The most complicated modules obtained by fusing first-row modules with those from the first or second row are certain rank-$2$ staggered modules.  More specifically, these are extensions of \hwms{} by other \hwms{}, meaning that they are (partially) defined by a short exact sequence,
\begin{equation*}
\dses{\IndMod{}}{\LogMod{}}{\IndMod{}'},
\end{equation*}
in which both $\IndMod{}$ and $\IndMod{}'$ are \hwms{}.  However, the decomposition of the fusion of second-row modules with other second-row modules has been shown to include extensions of \hwms{} by rank-$2$ staggered modules, for example,
\begin{equation*}
\dses{\LogMod{}}{\MegaMod{\brac{3}}}{\IrrMod{1,5}},
\end{equation*}
for an (unidentified) staggered module $\LogMod{}$.

This is interesting for many reasons.  The first is obviously that this provides further explicit examples of physically relevant indecomposable modules.  In fact, somewhat similar modules have already been discussed in \cite{EbeVir06} where it was noted that (in our terminology) fusing second-row modules from the \emph{standard} extended Kac table sometimes led to what they called rank-$3$ representations.  These are even more complicated than the modules which we have discussed, but we stress that they have not yet been associated with any physical observable (such as a crossing probability).  The key fundamental difference is the appearance of rank-$3$ Jordan cells for $L_0$ --- our modules $\MegaMod{\brac{3}}$ and $\MegaMod{\brac{3}}'$ only possess rank-$2$ cells.

An example should clarify this point.  If we had reason to include the module $\IndMod{2,3}$ in the spectrum of percolation, then we could fuse it with itself.  We would then obtain a module $\MegaMod{\brac{4}}$ (say) which may be described as an extension of a rank-$2$ staggered module by another such module:
\begin{equation*}
\dses{\LogMod{}}{\MegaMod{\brac{4}}}{\LogMod{}'}.
\end{equation*}
In practice, we are not able to identify these staggered module components completely (we can only see vanishing singular vectors to grade $7$), but we can say that $\LogMod{}$ has \hwsm{} $\IndMod{1,1}$ whereas that of $\LogMod{}'$ is $\IrrMod{1,5} = \IrrMod{3,1}$.  The first rank-$3$ cell occurs at grade $2$ and contains $L_{-2} \ket{0}$ and the two generators of $\LogMod{}'$.

There is an obvious resemblance to our module $\MegaMod{\brac{3}}$.  Indeed, modulo our inability to see singular vectors beyond grade $7$, we could even realise this rank-$3$ module as an extension of a dimension $2$ \hwm{} (of unknown character) by $\MegaMod{\brac{3}}$.  We believe it worth emphasising the observation that these modules possess a logarithmic structure relating a \hwsm{} (here $\IndMod{1,1}$) and a module for which \emph{both} the generating singular vectors vanish (an irreducible type III$_-$ module, in the language of \cite{FeiVer84}).  This would be impossible for a rank-$2$ staggered module --- the competing vanishing logarithmic singular vectors would lead to conflicting logarithmic couplings \cite{RidLog07}.  The presence of a third module is thus necessary in this situation to alleviate the conflict.

The ``higher indecomposables'' that we have found may then be viewed in a sense as the next rung on the complexity ladder after the rank-$2$ staggered modules.  They therefore form an important special case in any quest to understand indecomposable Virasoro modules.  One outstanding question which deserves further study is why we always seem to find irreducible type III$_-$ modules appearing in these more complicated cases.  Na\"{\i}vely, one can identify the set of modules $\IndMod{r,s}$ from which the indecomposables are expected to be constructed (that is, using the standard $\func{\alg{sl}}{2}$-type rules).  However, with each higher-indecomposable case, we observe that one of the expected vanishing singular vectors has \emph{migrated} to a different module (yielding the irreducible type III$_-$ module, in particular).  This migration principle requires further clarification if we are to understand the fusion rings involving higher indecomposables.

A second reason to be interested in these new indecomposables relates to the notion of ``projective cover'' which has recently started to permeate the physics literature.  This terminology describes a (minimal) projective indecomposable module which has a quotient module isomorphic to the module being covered.  The qualifier ``projective'' here refers to the property that this indecomposable cannot itself be realised as a (non-trivial) quotient module of a larger indecomposable.  The recent interest in these concepts seems to be rooted in the important r\^{o}le that they play in the representation theory of (finite-dimensional) simple Lie superalgebras \cite{ZouCat96,GerInd98,BruTil04} where they form an ideal in the representation ring (under tensor product).  Something very similar appears to happen in the fusion ring of our \lcft{}.  For example, the first-row indecomposables $\IndMod{1,3k + \ell}$ with $\ell=1,2$ are naturally realised as quotients of the larger indecomposables $\LogMod{1,3k + \ell}$.  The latter have not been observed to occur as quotients of still larger indecomposables, and moreover close (together with the irreducibles $\IrrMod{1,3k}$) under the fusion product.  It is therefore very natural to suspect that the $\IrrMod{1,3k}$ and the $\LogMod{1,3k + \ell}$ are projective (in some category), hence that the $\LogMod{1,3k + \ell}$ are projective covers of the $\IndMod{1,3k + \ell}$ and the $\IrrMod{1,3k + \ell}$.

Extending this speculation to the second row, we ask ourselves if the modules $\LogMod{2,3 \brac{k+1/2} + \ell}$ with $\ell=1,2$ and $\IndMod{2,3 \brac{k+1/2}}$ can be projective (in some sense).  Whilst the fusion with the first-row modules seems to close on this set of indecomposables, we have observed that fusing second-row modules with other second-row modules generates new indecomposables.  This contradicts the behaviour one expects from projective modules under fusion, so we are forced to conclude that the second-row modules are not projective (in the sense we want).  But this should not be surprising, as we have already noted that the second-row modules we have discovered are physically unsatisfactory as they contain non-vanishing singular vectors which are orthogonal to the entire module.  Physically, this leads to an unacceptable degeneracy in the corresponding $2$-point functions.

We concluded at the end of \secref{secWatts} that this means that there should be a larger indecomposable covering of the $\LogMod{2,3 \brac{k+1/2} + \ell}$ on which the inner product is non-degenerate.  We can now speculate that these sought-for larger coverings may just be the projective covers of $\LogMod{2,3 \brac{k+1/2} + \ell}$ in some physically appropriate category (in which such covers exist).  A thorough understanding the nature of these covers (and categories) remains the fundamental outstanding problem at this point, for such knowledge should be invaluable in describing the fusion ring in generality.  Indeed, it seems likely that such knowledge will be indispensable, as the ``brute-force'' methods currently employed to compute fusion rules have rather modest upper-bounds on the module complexities which can be handled.

We add however that the process by which we have so far investigated this covering phenomenon --- explicitly computing fusion rules --- is subject to the following observation.  We fuse a (physically unsatisfactory) module with other modules until we find a larger (hopefully physically satisfactory) module in which the original unsatisfactory module is realised as a \emph{submodule}.  Mathematically, this suggests that the cover we should be trying to understand is not the projective cover, but the ``injective cover'' (usually called the injective hull or envelope)\footnote{An injective module is one which cannot be realised as a proper \emph{submodule} of an indecomposable.} \cite{LamLec99}.  However, it should also be clear that the fundamental physical requirement is that the module carry a non-degenerate inner product (more precisely that the corresponding set of $2$-point functions are non-degenerate).  In other words, we want our modules to be isomorphic to their restricted duals\footnote{More generally, we want the restricted dual of each module to also be present in the spectrum (this is just the requirement that each field has a unique conjugate field).  This is not automatically satisfied.  For example, the \hwm{} $\IndMod{1,1}$ has a restricted dual which is not a \hwm{}, but is indecomposable, with maximal \hwsm{} isomorphic to the irreducible vacuum module $\IrrMod{1,1}$.  (This indecomposable module has no logarithmic structure however.)  But, this restricted dual cannot be included in the spectrum, as an irreducible vacuum module precludes the presence of any other modules.}.  This suggests that what we should be searching for are (minimal) self-dual indecomposables in which our physically unsatisfactory modules are realised as submodules (or as quotients).

Let us make one small observation however.  In \figref{figMoreDecomps} we have illustrated two particular fusion decompositions involving second-row modules.  We draw attention to the indecomposable module we have called $\LogMod{1/3}$, specifically to the fact that it has a submodule isomorphic to $\IndMod{1,3}$.  The latter is therefore not injective in any category containing $\LogMod{1/3}$ (which is physically required by fusion if $\IndMod{1,2}$ and $\IndMod{2,5/2}$ are present).  But, $\IndMod{1,3} = \IrrMod{1,3}$ is irreducible, so it is self-dual (its inner product is non-degenerate), and it is easy to see that under this condition, non-injectivity is equivalent to non-projectivity.  This contradicts our original heuristic notion that because the first-row modules $\IrrMod{1,3k}$ and $\LogMod{1,3k + \ell}$ ($\ell=1,2$) close under fusion, they must be projective \emph{in some sense}.  This counterexample serves to remind us that none of this is mathematically precise, and requiring projectivity in the mathematical sense is very likely too strong.  We would like to emphasise once again that projectivity (and injectivity) do not seem to have any obvious physical interpretation, and instead it is self-duality which is distinguished in this regard.

A third reason to be interested in the results presented here is the observation that fusing our second-row modules with themselves produces a vacuum module $\MegaMod{\brac{3}}$ which looks nothing like the vacuum module $\LogMod{1,5}$ obtained from the first-row modules.  It is immediately apparent from \figsref{figI1s}{figWxWDecomp} that $\LogMod{1,5}$ is neither a submodule nor a quotient module of $\MegaMod{\brac{3}}$.  More importantly, the vacuum of $\LogMod{1,5}$ has non-zero norm (unity by convention) whereas that of $\MegaMod{\brac{3}}$ is necessarily null (\eqnref{eqnNullVac}).  This proves that these vacua cannot be identified, not even if we proposed some even-larger indecomposable vacuum module covering both $\LogMod{1,5}$ and $\MegaMod{\brac{3}}$.  Indeed, it was proven in \cite[App.~C]{RidLog07} that a null first-row vacuum leads to an identically vanishing expression for Cardy's horizontal crossing probability.

We therefore seem to have two distinct vacua.  This is problematic in a chiral \cft{}, as it is customary to take the uniqueness of the vacuum as axiomatic.  The reason for this is that distinct vacua give rise to distinct energy-momentum tensors, hence distinct Virasoro algebras (in general, distinct chiral algebras).  The theory then decomposes\footnote{Is it possible for the two distinct vacua to give rise to distinct chiral algebras which are coupled together via some sort of indecomposable structure?  This is certainly not the case here, but it is nevertheless interesting to consider such a possibility.} into the direct sum of two (physically non-interacting) \cfts{}.

This logic forces us to take issue with the assumption that we can express percolation in terms of a \emph{chiral} \cft{}.  Although this is customary in the literature, the correct structure is that of (the boundary sector of) a \bcft{}.  This is something more than a chiral theory as each field, interpreted as a boundary condition changing operator, comes equipped with two labels describing which boundary condition it changes and what it changes into.  A pair of boundary labels is said to define a given boundary \emph{sector}, and the set of fields equipped with these labels defines the spectrum of this boundary sector.  Unfortunately, the allowed (conformal) boundary conditions of critical percolation are not particularly well understood at this point (see \cite{RasWEx08} for some recent work in this direction), and this is almost surely the reason why much of the literature has traditionally ignored the boundary labels that should appear on these fields.

In any case, it is clear that each distinct boundary label $\alpha$ must admit a distinct identity field of the form $\mathbf{1}^{\alpha \alpha}$ (which does not change the boundary condition), hence a distinct vacuum $\ket{0}^{\alpha \alpha}$ and vacuum module.  Our two vacuum modules $\LogMod{1,5}$ and $\MegaMod{\brac{3}}$ are therefore not in contradiction with the axioms of \cft{} if they can be identified as the vacuum modules of different boundary sectors.  This is a very natural conclusion as the boundary conditions invoked by Cardy and Watts in their respective seminal works are not the same.

Finally, we would like to briefly revisit the problem of inconsistencies in \lcfts{}.  As mentioned at the end of \secref{secLM23}, the conformal invariance of the vacuum prevents certain combinations of modules from appearing together in chiral theories.  In particular, the modules $\LogMod{1,5}$ and $\LogMod{3,1}$ are mutually incompatible in this respect.  We emphasise however that we have found \emph{no} such consistency problems in our investigation of Watts' (and Cardy's) crossing probability, which is (subjectively perhaps) strong evidence for the validity of our work.  There are nevertheless several means by which one can try to argue inconsistencies away, with varying degrees of success, and studying these arguments is important for determining whether percolation can admit still more general fields.  For example, we could try setting all troublesome correlators to zero.  Unfortunately, this does not alleviate the problem, as the invariant-vacuum \pdes{} at the root of the inconsistencies are not homogeneous (so do not admit zero as a solution).  A second attempt might be to note that since two mutually incompatible modules must share a common submodule, it would be enough to postulate that the common module actually appears with multiplicity two.  However, in the case of $\LogMod{1,5}$ and $\LogMod{3,1}$ the common submodule is actually the vacuum module $\IndMod{1,1}$, so this postulate runs afoul of the unique vacuum axiom of chiral \cft{}.

A better loophole to exploit is therefore the simple fact that we are working within a \bcft{}, so there are many different sectors to consider.  Mathematically, the conclusion of the inconsistency argument in this framework becomes the following:  If $\LogMod{3,1}$ does indeed have a physical interpretation in critical percolation, then it can only appear in boundary sectors whose boundary labels are \emph{disjoint} from those of $\LogMod{1,5}$.  In this way, one is prevented from forming the troublesome $2$-point correlation function of the corresponding fields $\func{\lambda_{1,5}}{z}$ and $\func{\lambda_{3,1}}{w}$.  Note that we are not setting these correlators to zero here, rather we are denying their existence (meaning) altogether.

It therefore follows that a physically convincing interpretation for $\lambda_{3,1}$ (or $\phi_{2,1}$) as a boundary field in critical percolation is not completely ruled out!  If such an interpretation is found then one must ask whether there is a sequence of boundary fields which (when applied in the correct order) changes a boundary condition associated with $\lambda_{3,1}$ into a boundary condition associated with $\lambda_{1,5}$.  If there is, then we have a (possibly consistent) mechanism whereby both fields can coexist within the theory.  If there is no such sequence, then the boundary sectors decompose into (at least) two ``connected'' components, and we are effectively left with two different \bcfts, distinguished by the class of boundary conditions we can impose.

We conclude by mentioning that it may be possible to (objectively) test the above conclusion using stochastic Loewner evolution.  There, one has an interpretation in percolation for $\phi_{1,2}$ as the trace generator for $\kappa = 6$ and for $\phi_{2,1}$ as the trace generator for $\kappa = 8/3$ (but no longer in percolation!).  There are therefore ``composite'' interpretations for both $\LogMod{1,5}$ and $\LogMod{3,1}$ via fusion.  It is an open question at present if there is any obstacle to considering a system with both values of $\kappa$, such that both values interact.  One subtlety worth making explicit is that in \cft{} we generally do not work with the boundary, but rather employ the method of images to recover a theory without boundary.  This is necessary to obtain a conformally invariant vacuum, the object at the heart of the inconsistency derivations.  This method of images does not seem to play a r\^{o}le in stochastic Loewner evolution, so it is not clear that a conformally invariant vacuum state is even present in the latter approach.  Finally, one has to wonder whether considering evolutions for traces with two different values of $\kappa$ might not correspond to \cfts{} with some sort of defect wall corresponding to the discontinuity in $\kappa$.  In any case, it should be clear that there are still many important fundamental questions to be resolved, and we hope to report on these at a later date.

\section*{Acknowledgements}

I would like to thank Pierre Mathieu for discussions during the formative stages of this project, so long ago, and Matthias Gaberdiel for his encouragement and suggestions when things were floundering and morale was low.  I would like to especially thank Jake Simmons for explaining his work to me and for his hospitality in Oxford.  I also thank John Cardy, B\'{e}njamin Doyon, Kalle Kyt\"{o}l\"{a}, Thomas Quella and Hubert Saleur for interesting (and inspiring) discussions on logarithmic structure, boundary conditions and SLE.


\begin{thebibliography}{10}

\bibitem{KesPer82}
H~Kesten.
\newblock {\em {Percolation Theory for Mathematicians}}, volume~2 of {\em
  Progress in Probability and Statistics}.
\newblock Birkh\"{a}user, Boston, 1982.

\bibitem{GriPer89}
G~Grimmett.
\newblock {\em {Percolation}}, volume 321 of {\em Grundlehren der
  Mathematischen Wissenschaften}.
\newblock Springer-Verlag, Berlin, 1999.

\bibitem{LanCon94}
R~Langlands, P~Pouliot, and Y~Saint-Aubin.
\newblock {Conformal Invariance in Two-Dimensional Percolation}.
\newblock {\em Bull. Amer. Math. Soc.}, 30:1--61, 1994.

\bibitem{CarLec01}
J~Cardy.
\newblock {Lectures on Conformal Invariance and Percolation}.
\newblock \texttt{arXiv:math-ph/0103018}.

\bibitem{CarCri92}
J~Cardy.
\newblock {Critical Percolation in Finite Geometries}.
\newblock {\em J. Phys.}, A25:L201--L206, 1992.
\newblock \texttt{arXiv:hep-th/9111026}.

\bibitem{DiFCon97}
P~Di Francesco, P~Mathieu, and D~S\'{e}n\'{e}chal.
\newblock {\em {Conformal Field Theory}}.
\newblock Graduate Texts in Contemporary Physics. Springer-Verlag, New York,
  1997.

\bibitem{LanUni92}
R~Langlands, C~Pichet, P~Pouliot, and Y~Saint-Aubin.
\newblock {On the Universality of Crossing Probabilities in Two-Dimensional
  Percolation}.
\newblock {\em J. Stat. Phys.}, 67:553--574, 1992.

\bibitem{SmiCri01}
S~Smirnov.
\newblock {Critical Percolation in the Plane: Conformal Invariance, Cardy's
  Formula, Scaling Limits}.
\newblock {\em C. R. Acad. Sci. Paris S\'{e}r. I Math.}, 333:239--244, 2001.

\bibitem{LawValI01}
G~Lawler, O~Schramm, and W~Werner.
\newblock {Values of Brownian Intersection Exponents, I: Half-Plane Exponents}.
\newblock {\em Acta Math.}, 187:237--273, 2001.
\newblock \texttt{arXiv:math.PR/9911084}.

\bibitem{SchSca00}
O~Schramm.
\newblock {Scaling Limits of Loop-Erased Random Walks and Uniform Spanning
  Trees}.
\newblock {\em Isr. J. Math.}, 118:221--288, 2000.
\newblock \texttt{arXiv:math.PR/9904022}.

\bibitem{BauCon04}
M~Bauer and D~Bernard.
\newblock {Conformal Transformations and the SLE Partition Function
  Martingale}.
\newblock {\em Ann. Henri Poincar\'{e}}, 5:289--326, 2004.
\newblock \texttt{arXiv:hep-th/0305061}.

\bibitem{KytFro08}
K~Kyt\"{o}l\"{a}.
\newblock {From SLE to the Operator Content of Percolation}.
\newblock \texttt{arXiv:0804.2612 [math-ph]}.

\bibitem{CarLog99}
J~Cardy.
\newblock {Logarithmic Correlations in Quenched Random Magnets and Polymers}.
\newblock \texttt{arXiv:cond-mat/9911024}.

\bibitem{FloNot06}
M~Flohr and A~Muller-Lohmann.
\newblock {Notes on Non-Trivial and Logarithmic CFTs with $c=0$}.
\newblock {\em J. Stat. Mech.}, 0604:002, 2006.
\newblock \texttt{arXiv:hep-th/0510096}.

\bibitem{FeiAnn92}
B~Feigin, T~Nakanishi, and H~Ooguri.
\newblock {The Annihilating Ideals of Minimal Models}.
\newblock {\em Int. J. Mod. Phys.}, A7:217--238, 1992.

\bibitem{RidPer07}
P~Mathieu and D~Ridout.
\newblock {From Percolation to Logarithmic Conformal Field Theory}.
\newblock {\em Phys. Lett.}, B657:120--129, 2007.
\newblock \texttt{arXiv:0708.0802 [hep-th]}.

\bibitem{RozQua92}
L~Rozansky and H~Saleur.
\newblock {Quantum Field Theory for the Multivariable Alexander-Conway
  Polynomial}.
\newblock {\em Nucl. Phys.}, B376:461--509, 1992.

\bibitem{GurLog93}
V~Gurarie.
\newblock {Logarithmic Operators in Conformal Field Theory}.
\newblock {\em Nucl. Phys.}, B410:535--549, 1993.
\newblock \texttt{arXiv:hep-th/9303160}.

\bibitem{GurCon02}
V~Gurarie and A~Ludwig.
\newblock {Conformal Algebras of 2-D Disordered Systems}.
\newblock {\em J. Phys.}, A35:L377--L384, 2002.
\newblock \texttt{arXiv:cond-mat/9911392}.

\bibitem{CarStr02}
J~Cardy.
\newblock {The Stress Tensor in Quenched Random Systems}.
\newblock In A~Cappelli and G~Mussardo, editors, {\em Statistical Field
  Theories}, volume~73 of {\em NATO Science Series II: Mathematics, Physics and
  Chemistry}, pages 215--222. Kluwer, Dordrecht, 2002.
\newblock \texttt{arXiv:cond-mat/0111031}.

\bibitem{KogStr02}
I~Kogan and A~Nichols.
\newblock {Stress Energy Tensor in LCFT and the Logarithmic Sugawara
  Construction}.
\newblock {\em JHEP}, 0201:029, 2002.
\newblock \texttt{arXiv:hep-th/0112008}.

\bibitem{GurCon04}
V~Gurarie and A~Ludwig.
\newblock {Conformal Field Theory at Central Charge $c=0$ and Two-Dimensional
  Critical Systems with Quenched Disorder}.
\newblock In M~Shifman, editor, {\em From Fields to Strings: Circumnavigating
  Theoretical Physics. Ian Kogan Memorial Collection}, volume~2, pages
  1384--1440. World Scientific, Singapore, 2005.
\newblock \texttt{arXiv:hep-th/0409105}.

\bibitem{RidMin07}
P~Mathieu and D~Ridout.
\newblock {The Extended Algebra of the Minimal Models}.
\newblock {\em Nucl. Phys.}, B776:365--404, 2007.
\newblock \texttt{arXiv:hep-th/0701250}.

\bibitem{ZamInf85}
A~Zamolodchikov.
\newblock {Infinite Additional Symmetries in Two-Dimensional Conformal Quantum
  Field Theory}.
\newblock {\em Theor. Math. Phys.}, 65:1205--1213, 1985.

\bibitem{CauLog97}
J~Caux, I~Kogan, A~Lewis, and A~Tsvelik.
\newblock {Logarithmic Operators and Dynamical Extension of the Symmetry Group
  in the Bosonic $SU \left( 2 \right)_0$ and SUSY $SU \left(2 \right)_2$ WZNW
  Models}.
\newblock {\em Nucl. Phys.}, B489:469--484, 1997.
\newblock \texttt{arXiv:hep-th/9606138}.

\bibitem{NicLog01}
A~Nichols.
\newblock {Logarithmic Currents in the $SU \left( 2 \right)_0$ WZNW Model}.
\newblock {\em Phys. Lett.}, B516:439--445, 2001.
\newblock \texttt{arXiv:hep-th/0102156}.

\bibitem{BerQua01}
D~Bernard and A~LeClair.
\newblock {Quasi-Spin-Charge Separation and the Spin Quantum Hall Effect}.
\newblock {\em Phys. Rev.}, B64:045306, 2001.
\newblock \texttt{arXiv:cond-mat/0003075}.

\bibitem{LudFre00}
A~Ludwig.
\newblock {A Free Field Representation of the $Osp \left( 2 \mid 2 \right)$
  Current Algebra at Level $k = -2$, and Dirac Fermions in a Random $SU \left(
  2 \right)$ Gauge Potential}, 2000.
\newblock \texttt{arXiv:cond-mat/0012189}.

\bibitem{SalGL106}
H~Saleur and V~Schomerus.
\newblock {The $GL \left( 1 \mid 1 \right)$ WZW Model: From Supergeometry to
  Logarithmic CFT}.
\newblock {\em Nucl. Phys.}, B734:221--245, 2006.
\newblock \texttt{arXiv:hep-th/0510032}.

\bibitem{SalSU207}
H~Saleur and V~Schomerus.
\newblock {On the $SU \left( 2 \mid 1 \right)$ WZW Model and its Statistical
  Mechanics Applications}.
\newblock {\em Nucl. Phys.}, B775:312--340, 2007.
\newblock \texttt{arXiv:hep-th/0611147}.

\bibitem{CreBra08}
T~Creutzig, T~Quella, and V~Schomerus.
\newblock {Branes in the $GL \left( 1 \mid 1 \right)$ WZNW Model}.
\newblock {\em Nucl. Phys.}, B792:257--283, 2008.
\newblock \texttt{arXiv:0708.0853 [hep-th]}.

\bibitem{FjeLog02}
J~Fjelstad, J~Fuchs, S~Hwang, A~Semikhatov, and I~Yu Tipunin.
\newblock {Logarithmic Conformal Field Theories via Logarithmic Deformations}.
\newblock {\em Nucl. Phys.}, B633:379--413, 2002.
\newblock \texttt{arXiv:hep-th/0201091}.

\bibitem{FeiMod06}
B~Feigin, A~Gainutdinov, A~Semikhatov, and I~Yu Tipunin.
\newblock {Modular Group Representations and Fusion in Logarithmic Conformal
  Field Theories and in the Quantum Group Center}.
\newblock {\em Comm. Math. Phys.}, 065:47--93, 2006.
\newblock \texttt{arXiv:hep-th/0504093}.

\bibitem{FeiLog06}
B~Feigin, A~Gainutdinov, A~Semikhatov, and I~Yu Tipunin.
\newblock {Logarithmic Extensions of Minimal Models: Characters and Modular
  Transformations}.
\newblock {\em Nucl. Phys.}, B757:303--343, 2006.
\newblock \texttt{arXiv:hep-th/0606196}.

\bibitem{ReaExa01}
N~Read and H~Saleur.
\newblock {Exact Spectra of Conformal Supersymmetric Nonlinear Sigma Models in
  Two Dimensions}.
\newblock {\em Nucl. Phys.}, B613:409--444, 2001.
\newblock \texttt{arXiv:hep-th/0106124}.

\bibitem{PeaLog06}
P~Pearce, J~Rasmussen, and J-B Zuber.
\newblock {Logarithmic Minimal Models}.
\newblock {\em J. Stat. Mech.}, 0611:017, 2006.
\newblock \texttt{arXiv:hep-th/0607232}.

\bibitem{ReaAss07}
N~Read and H~Saleur.
\newblock {Associative-Algebraic Approach to Logarithmic Conformal Field
  Theories}.
\newblock {\em Nucl. Phys.}, B777:316--351, 2007.
\newblock \texttt{arXiv:hep-th/0701117}.

\bibitem{RasFus07}
J~Rasmussen and P~Pearce.
\newblock {Fusion Algebra of Critical Percolation}.
\newblock {\em J. Stat. Mech.}, 0709:002, 2007.
\newblock \texttt{arXiv:0706.2716 [hep-th]}.

\bibitem{RasWEx08}
J~Rasmussen and P~Pearce.
\newblock {W-Extended Fusion Algebra of Critical Percolation}, 2008.
\newblock \texttt{arXiv:0804.4335 [hep-th]}.

\bibitem{EbeVir06}
H~Eberle and M~Flohr.
\newblock {Virasoro Representations and Fusion for General Augmented Minimal
  Models}.
\newblock {\em J. Phys.}, A39:15245--15286, 2006.
\newblock \texttt{arXiv:hep-th/0604097}.

\bibitem{SimPer07}
J~Simmons, P~Kleban, and R~Ziff.
\newblock {Percolation Crossing Formulas and Conformal Field Theory}.
\newblock {\em J. Phys.}, A40:F771--784, 2007.
\newblock \texttt{arXiv:0705.1933 [cond-mat.stat-mech]}.

\bibitem{RidLog07}
P~Mathieu and D~Ridout.
\newblock {Logarithmic $M \left( 2,p \right)$ Minimal Models, their Logarithmic
  Couplings, and Duality}.
\newblock {\em Nucl. Phys.}, B801:268--295, 2008.
\newblock \texttt{arXiv:0711.3541 [hep-th]}.

\bibitem{GabLoc99}
M~Gaberdiel and H~Kausch.
\newblock {A Local Logarithmic Conformal Field Theory}.
\newblock {\em Nucl. Phys.}, B538:631--658, 1999.
\newblock \texttt{arXiv:hep-th/9807091}.

\bibitem{CarEff86}
J~Cardy.
\newblock {Effect of Boundary Conditions on the Operator Content of
  Two-Dimensional Conformally Invariant Theories}.
\newblock {\em Nucl. Phys.}, B275:200--218, 1986.

\bibitem{RunBou99}
I~Runkel.
\newblock {Boundary Structure Constants for the A-Series Virasoro Minimal
  Models}.
\newblock {\em Nucl. Phys.}, B549:563--578, 1999.
\newblock \texttt{arXiv:hep-th/9811178}.

\bibitem{FucTFTI02}
J~Fuchs, I~Runkel, and C~Schweigert.
\newblock {TFT Construction of RCFT Correlators 1. Partition Functions}.
\newblock {\em Nucl. Phys.}, B646:353--497, 2002.
\newblock \texttt{arXiv:hep-th/0204148}.

\bibitem{FjeUni06}
J~Fjelstad, J~Fuchs, I~Runkel, and C~Schweigert.
\newblock {Uniqueness of Open/Closed Rational CFT with given Algebra of Open
  States}.
\newblock \texttt{arXiv:hep-th/0612306}.

\bibitem{GabFro08}
M~Gaberdiel and I~Runkel.
\newblock {From Boundary to Bulk in Logarithmic CFT}.
\newblock {\em J. Phys.}, A41:075402, 2008.
\newblock \texttt{arXiv:0707.0388 [hep-th]}.

\bibitem{WatCro96}
G~Watts.
\newblock {A Crossing Probability for Critical Percolation in Two-Dimensions}.
\newblock {\em J. Phys.}, A29:L363--L368, 1996.
\newblock \texttt{arXiv:cond-mat/9603167}.

\bibitem{FeiVer84}
B~Feigin and D~Fuchs.
\newblock {Verma Modules over the Virasoro Algebra}.
\newblock In {\em Topology}, volume 1060 of {\em Lecture Notes in Mathematics},
  pages 230--245. Springer, Berlin, 1984.

\bibitem{GabFus94}
M~Gaberdiel.
\newblock {Fusion in Conformal Field Theory as the Tensor Product of the
  Symmetry Algebra}.
\newblock {\em Int. J. Mod. Phys.}, A9:4619--4636, 1994.
\newblock \texttt{arXiv:hep-th/9307183}.

\bibitem{NahQua94}
W~Nahm.
\newblock {Quasirational Fusion Products}.
\newblock {\em Int. J. Mod. Phys.}, B8:3693--3702, 1994.
\newblock \texttt{arXiv:hep-th/9402039}.

\bibitem{GabInd96}
M~Gaberdiel and H~Kausch.
\newblock {Indecomposable Fusion Products}.
\newblock {\em Nucl. Phys.}, B477:293--318, 1996.
\newblock \texttt{arXiv:hep-th/9604026}.

\bibitem{RohRed96}
F~Rohsiepe.
\newblock {On Reducible but Indecomposable Representations of the Virasoro
  Algebra}.
\newblock \texttt{arXiv:hep-th/9611160}.

\bibitem{BogInd74}
J~Bogn\'{a}r.
\newblock {\em {Indefinite Inner Product Spaces}}, volume~78 of {\em Ergebnisse
  der Mathematik und ihrer Grenzgebiete}.
\newblock Springer-Verlag, Berlin, 1974.

\bibitem{MooLie95}
R~Moody and A~Pianzola.
\newblock {\em {Lie Algebras with Triangular Decompositions}}.
\newblock Canadian Mathematical Society Series of Monographs and Advanced
  Texts. Wiley, New York, 1995.

\bibitem{DotCon84}
V~Dotsenko and V~Fateev.
\newblock {Conformal Algebra and Multipoint Correlation Functions in 2D
  Statistical Models}.
\newblock {\em Nucl. Phys.}, B240:312--348, 1984.

\bibitem{ChiInt92}
L~Chim and A~Zamolodchikov.
\newblock {Integrable Field Theory of the $q$-State Potts Model with $0<q<4$}.
\newblock {\em Int. J. Mod. Phys.}, A7:5317--5335, 1992.

\bibitem{DubExc06}
J~Dub\'{e}dat.
\newblock {Excursion Decompositions for SLE and Watts' Crossing Formula}.
\newblock {\em Prob. Theory Rel. Fields}, 134:453--488, 2006.
\newblock \texttt{arXiv:math.PR/0405074}.

\bibitem{ZouCat96}
Y~Zou.
\newblock {Categories of Finite Dimensional Weight Modules over Type I
  Classical Lie Superalgebras}.
\newblock {\em J. Alg.}, 180:459--482, 1996.

\bibitem{GerInd98}
J~Germoni.
\newblock {Indecomposable Representations of Special Linear Lie Superalgebras}.
\newblock {\em J. Alg.}, 209:367--401, 1998.

\bibitem{BruTil04}
J~Brundan.
\newblock {Tilting Modules for Lie Superalgebras}.
\newblock {\em Comm. Alg.}, 32:2251--2268, 2004.
\newblock \texttt{arXiv:math.RT/0209235}.

\bibitem{LamLec99}
T-Y Lam.
\newblock {\em {Lectures on Modules and Rings}}, volume 189 of {\em Graduate
  Texts in Mathematics}.
\newblock Springer-Verlag, Berlin, 1999.

\end{thebibliography}
\end{document}